\documentclass[preprint,showpacs,superscriptaddress,longbibliography]{revtex4-2}
\usepackage{amsmath,amssymb,bm}
\usepackage{mathrsfs}
\usepackage{textcomp}
\usepackage{commath}
\usepackage{mathtools}

\usepackage{hyperref}
\hypersetup{
    colorlinks=true,
    linkcolor=blue,
}

\usepackage{natbib}
\usepackage{placeins}
\usepackage[final]{graphicx} 
\usepackage{rotating}
\usepackage{epstopdf}
\usepackage{float}
\usepackage{amsmath}
\usepackage[lofdepth=1,lotdepth]{subfig}
\usepackage[usenames]{color} 

\usepackage{cancel} 

\usepackage[normalem]{ulem}
\renewcommand{\sout}[1]{} 

\newcommand{\add}[1]{\textcolor{black}{#1}}

\newcommand{\Ca}{\mathrm{Ca}}

\newcommand{\eg}{\mathrm{E_G}}

\graphicspath{{figures/}}  

\makeatletter

\newcommand{\numtoRoman}[1]{\expandafter\@slowromancap\romannumeral #1@}
\makeatother

\begin{document}

\title{Free-space and near-wall dynamics of a flexible sheet sedimenting in Stokes flow}
\author{Yijiang Yu}
\affiliation{
Department of Chemical and Biological Engineering\\
University of Wisconsin-Madison, Madison, WI 53706-1691
}
\author{Michael D. Graham}\email{Corresponding author. E-mail: mdgraham@wisc.edu}
\affiliation{
Department of Chemical and Biological Engineering\\
University of Wisconsin-Madison, Madison, WI 53706-1691
}
\date{\today}

\begin{abstract}
We present a numerical study of a thin elastic sheet with small extensibility freely sedimenting in a viscous fluid. Two scenarios are investigated: sedimentation in free space and near an infinite wall, where the wall may be vertical or tilted. Elastic sheets with a rest shape of a square are modeled with a finite-element-based continuum model that accounts for in-plane stretching and out-of-plane bending. The fluid motion is computed by the method of regularized Stokeslets in free space and regularized Blakelets near a wall. 
During sedimentation, the interplay between gravity and the elastic response of sheets gives rise to complex deformation and reorientation dynamics, measured by a dimensionless elasto-gravitational number ($\eg$).
In free space, sheets attain a stable orientation by aligning perpendicular to gravity. Sheets with larger deformability adopt more compact conformations and experience smaller hydrodynamic drag, thereby sedimenting faster. A sheet with a random initial orientation reorients to align perpendicular to gravity, accompanied by lateral drift due to the symmetry-breaking in conformations. We identified two reorientation mechanisms depending on flexibility.
When a sheet is placed near an infinite wall, sedimentation is hindered compared to that in free space due to wall-induced hydrodynamic drag. Near a vertical wall, sheets exhibit asymmetric conformations that cause the sheet to drift, with the drifting dynamics determined by $\eg$. The difference in flexibility leads to a non-monotonic trend in the evolution of wall-normal distance.
Near a tilted wall, sheets show qualitatively different dynamics when the wall angle is large: they either deposit on or slide along the wall with a fixed wall-normal distance. 

\end{abstract}
\maketitle

\section{INTRODUCTION} \label{sec:introduction}

The sedimentation dynamics of particles in a viscous fluid under gravitational or centrifugal forces have been receiving significant interest for their ubiquity in many industrial applications and biological systems~\cite{moths2013full,bjoornmalm2016dynamic,saintillan2005smooth,gruziel2018periodic,vologodskii1998sedimentation,weber2013sedimentation, backes2016preparation, xu2018liquid, zamani2021ultralight, hernandez2008high, wei2014size, das2018carbon, zhang2018large}.
Extensive research has been performed to explore sedimentation dynamics for various particle configurations, such as filaments~\cite{shashank2023dynamics,bukowicki2019sedimenting,jianzhong2003effects,li2013sedimentation,lagomarsino2005hydrodynamic,saggiorato2015conformations,cunha2022settling}, rings~\cite{gruziel2019stokesian,gruziel2018periodic},  cylinders~\cite{makino2003sedimentation,chajwa2019kepler,salez2015elastohydrodynamics,saintyves2016self, rallabandi2017rotation}, and spheroids~\cite{mitchell2015sedimentation,peltomaki2013sedimentation,matsunaga2016reorientation}. However, one important geometry has rarely been investigated: thin elastic sheets. Sedimentation of elastic sheets is prevalent in the synthesis and processing of polymer films, nanosheets, and quantum dots~\cite{das2018carbon, backes2016preparation, xu2018liquid, zamani2021ultralight, hernandez2008high, wei2014size, zhang2018large}. For example, graphene flakes obtained by liquid-phase exfoliation are separated via centrifugation~\cite{backes2016preparation, xu2018liquid, zamani2021ultralight, hernandez2008high}. Therefore, the knowledge of sheet-like particle sedimentation is crucial to predict sheet conformations obtained from centrifugation. In this paper, we aim to provide a fundamental understanding of thin sheets sedimenting in a viscous fluid. Sedimentation dynamics are affected by many factors, including particle stiffness~\cite{li2013sedimentation,cunha2022settling}, aspect ratio~\cite{jianzhong2003effects}, fluid properties~\cite{becker1996sedimentation, singh2000sedimentation}. In this work, we focus on particle bending stiffness and wall effects. Below, we review some related works of other particle configurations to provide insights for understanding elastic sheets. 

It is known that rigid particles, like spheroids and rods, sedimenting in an unbounded environment maintain their initial orientation due to Stokes flow reversibility~\cite{graham2018microhydrodynamics}. A tilted initial orientation can induce continuous drift of the particle in the lateral direction whereby the particle sediments with a fixed angle determined by initial orientation. In contrast, deformability allows particles to adjust their orientation and introduces more complex sedimentation dynamics. For instance, sedimenting flexible slender fibers and filaments form varied shapes in free space and have a unique and stable orientation with the end-to-end vector perpendicular to gravity, as a result of nonlocal hydrodynamic interactions~\cite{lagomarsino2005hydrodynamic}.
In a computational study, Li \textit{et al.} found that a non-Brownian weakly flexible filament with random initial conditions reorients to a unique orientation and conformation, where the reorientation trajectories are restricted to a cloud with envelope size determined by flexibility~\cite{li2013sedimentation}.
Cunha \textit{et al.} studied non-Brownian sedimenting flexible filaments modeled with a bead-spring system and characterized the final shapes of filaments varying from almost horizontal to horseshoe-like based on flexibility~\cite{cunha2022settling}. Moreover, they identified three different reorientation dynamics with decreasing resistance to gravity-induced deformation: rotating, bending, and snaking.
Gruziel and coworkers examined a knotted deformable closed chain with a rest shape of ring sedimenting in a viscous fluid. In both experiments and simulations, they observed the knots often attained a toroidal structure with oscillating intertwined loops during sedimentation~\cite{gruziel2018periodic}. In a later study, Gruziel \textit{et al.} also gave a detailed investigation of loop flexibility and found rich sedimentation dynamics~\cite{gruziel2019stokesian}.
For more complex geometries, Peltomaki and Gompper did a numerical study of deformable red blood cell sedimentation using multi-particle collision dynamics. They characterized and presented a phase diagram of three gravity-induced shapes depending on elasticity: parachutes, teardrops, and fin-tailed spheres~\cite{peltomaki2013sedimentation}. Matsunaga \textit{et al.} extended the study by considering red blood cells with different ratios of internal and external fluid viscosity to investigate the reorientation dynamics during sedimentation~\cite{matsunaga2016reorientation}.
The studies of single-particle sedimenting in free space benefit the understanding of more complex scenarios. When multiple particles sediment together, the deformation and collective sedimentation dynamics of particles are affected by inter-particle hydrodynamic interactions. Recent experiments and simulations have reported interesting hydrodynamic repulsion and attraction, as well as shape deformations~\cite{saggiorato2015conformations, shashank2023dynamics, bukowicki2019sedimenting, chajwa2019kepler}. We will not elaborate on this topic as the present work focuses on single-particle dynamics.

In contrast to sedimenting in free space, the sedimenting dynamics can be strongly modified with the presence of boundaries. It is known that particles next to a wall sediment slower compared to free space because the wall-induced drag hinders sedimentation. Rigid nonspherical particles such as a tilted rod close to a wall no longer maintain the orientation. Instead, they can rotate and migrate away from the wall. Russel \textit{et al.} first reported characteristic glancing or reversing dynamics of a rod sedimenting near a vertical wall, determined by the initial angle of the rod approaching the wall~\cite{russel1977rods}.
Mitchell and Spagnolie later extended the study to a rigid spheroidal particle with an arbitrary aspect ratio next to a tilted wall~\cite{mitchell2015sedimentation} at zero Reynolds number. They developed analytical solutions based on the method of images for randomly oriented prolate and oblate spheroids and found 3D dynamics like glancing, reversing, and wobbling based on the symmetry of initial orientations. In addition, they identified a stable sliding trajectory when the wall is slightly tilted.

Besides sedimentation in stationary fluid, recent works have explored the dynamics of elastic sheets in fundamental linear flows and their dependence on sheet elasticity. In shear flow, Xu and Green performed Brownian dynamics simulation and observed a cyclic crumple-stretch motion in a square sheet~\cite{xu2014brownian}. Silmore and coworkers addressed the influence of bending rigidity and initial orientation on the dynamics of a non-Brownian flexible sheet~\cite{silmore2021buckling}. Recently, Perrin and coworkers experimentally identified a decrease in the buckling threshold when two parallel rectangles tumbled in a shear cell due to hydrodynamic interactions between sheets~\cite{perrin2023hydrodynamic}. In extensional flow, Yu and Graham numerically investigated the coil-stretch-like transition for soft elastic sheets that show conformational hysteresis as a result of the interplay between flow strength and sheet elasticity~\cite{yu2021coil,yu2022wrinkling}. 

The aforementioned studies provide good insights into flexible particle sedimentation, and the similarity between elastic sheets and elastic filaments implies that rich dynamics can be found for a sedimenting 2D sheet in a viscous fluid. In this work, we present a numerical study to systematically investigate the sedimentation dynamics of an almost inextensible elastic sheet by addressing the influence of sheet bending elasticity and wall effects.
The structure of this work is as follows. In Section \ref{sec:sed_methods}, we present the model and numerical methods to simulate an almost inextensible sheet. We introduce two different flow solvers to account for the sedimentation in free space and near-wall sedimentation, separately. 
In Section \ref{sec:sed_results}, we introduce simulation results by first discussing the free space sedimentation, which is purely determined by the effect of sheet flexibility. Next, we place an elastic sheet next to an infinite solid wall to examine the wall effects, where the wall can be tilted.
We conclude in Section \ref{sec:sed_conclusion}.


\section{MODEL DEVELOPMENT} \label{sec:sed_methods}

\subsection{Model setup}

We consider an elastic thin sheet with a square rest shape that sediments in a stationary viscous fluid. The fluid is Newtonian with viscosity $\eta$ and density $\rho_f$. 
The square sheet has edge $2a$ and thickness $h$ such that $h \ll a$. Thus, the sheet has a traction-free edge, along which fluid exerts no forces. 
The weight of the sheet can be described as $4\rho_s g a^2h$, where $\rho_s$ is density and $g$ is the gravitational constant. The gravity, or the relative centrifugal force in the case of centrifugation, acts along the positive $x$ axis with $y$ and $z$ marked as lateral directions.
In this work, we investigate two scenarios: the sheet sediments 1) in an unbounded fluid, and 2) next to an infinite and solid wall, where the wall can be vertical or tilted. The initial condition of the sheet is described by an orientation angle $\phi$, which is defined as the angle between the normal vector $\mathbf{n}$ of the sheet and the positive $y$ axis. (i.e. $\phi = 90^\circ$ indicates a sheet oriented perpendicular to gravity and $\phi = 0^\circ$ when the sheet sediments parallel to gravity.)
\begin{figure}[!t]
    \centering
    \captionsetup{justification=raggedright}
\includegraphics[width=0.6\textwidth]{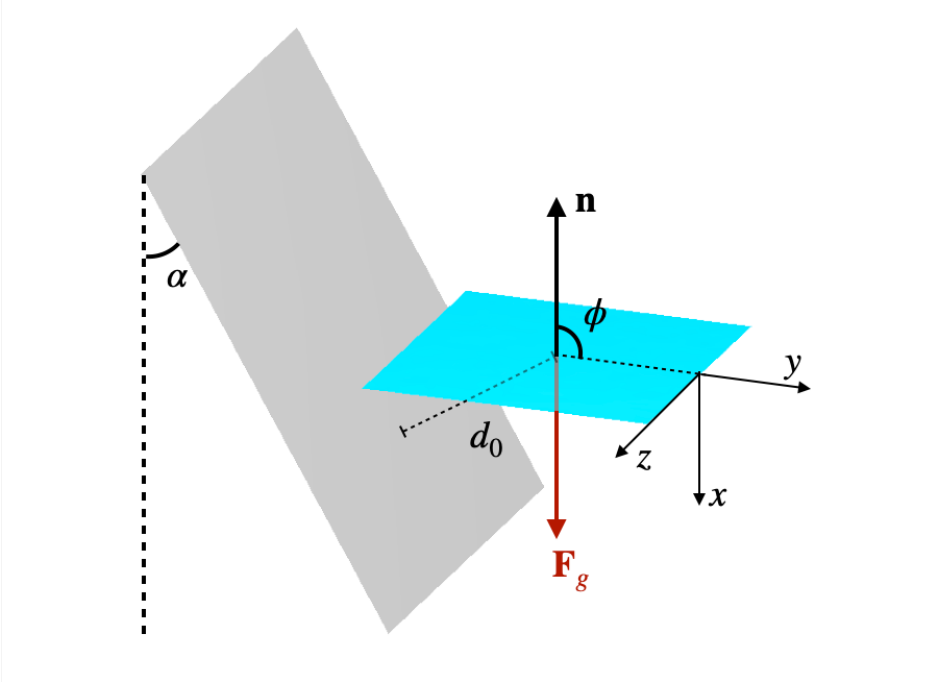}
    \caption{Illustration of a square sheet sedimenting next to a tilted wall. The case shows the sheet with initial wall-normal distance $d_0 = 1.5$ is oriented $\phi = 90^\circ$ and the titled wall angle is $\alpha = 30^\circ$.}
    \label{fig:model_setup}    
\end{figure} 
In the near-wall scenario, the wall can be tilted with an angle $\alpha$. If the wall is vertical ($\alpha = 0^\circ$), it aligns with the plane $y = 0$, such that $y$ becomes the wall-normal direction. The initial distance between the center of mass of the sheet and the wall surface is measured by the wall-normal distance $d_0$. 
Figure \ref{fig:model_setup} illustrates the setup by placing a sheet with $\phi = 90^\circ$ next to a wall tilted by $\alpha = 30^\circ$. \add{In simulations, we model the tilted wall by modifying the gravity direction rather than the wall angle. For instance, the case in Fig.~\ref{fig:model_setup} is simulated by tilting gravity with $30^\circ$ toward a vertical wall, and the initial orientation of the sheet is adjusted to maintain perpendicular to the modified gravity direction. For visualization purposes, the results presented for the titled wall scenario will follow the setup in Fig.~\ref{fig:model_setup}.}

We model the sheet as a continuum with a no-slip surface, so material points on the surface move with the same velocity as the local fluid.
The no-slip boundary condition also implies that the sheet is impermeable -- if the sheet moves with the local fluid velocity, then no fluid is passing through the sheet.
The sheet surface is discretized into triangular elements with a node at each corner. Detailed numerical methods will be introduced in the following section. 

The total strain energy of the elastic sheet can be written as the sum of in-plane  strain energy $E_s$ and out-of-plane bending energy $E_b$:
\begin{equation}
    E = E_s + E_b.
\end{equation}
Here, the in-plane strain energy is evaluated by integrating the strain energy density $W$ over the sheet surface. We choose a nonlinear neo-Hookean model (NH), as it is often used to model rubber-like polymer structures, with the form:

\begin{equation}
    W_\mathrm{NH} = \frac{G}{2}\left( I_1 -3\right) .
\end{equation}

The two-dimensional shear modulus $G$ scales with the equilibrium thickness $h$. The term $I_1$ is the first invariant of the right Cauchy-Green deformation tensor, and depends on the local principal stretch ratios $\lambda_i$ along the tangential direction of the sheet surface:

\begin{equation}
	I_{1}=\lambda_1^2+\lambda_2^2+\frac{1}{\lambda_1^2\lambda_2^2}.
\end{equation}
The last term denotes $\lambda_3$, as the stretch ratio along the thickness direction, is obtained by incompressibility of the material ($\lambda_1\lambda_2\lambda_3 = 1$). 

For the bending energy $E_b$, we apply a simple bending model that sums the energy due to dihedral angles $\theta_{\alpha\beta}$ between neighboring elements~\cite{Fedosov2010,fedosov2010multiscale}:
\begin{equation}
    E_b = \sum_{\mathrm{adj}\ \alpha,\beta}k_b \left(1-\cos(\theta_{\alpha\beta} - \theta_0)\right),
\label{eq:bending}
\end{equation}
where $k_b$ is the bending constant calculated from bending stiffness $K_B$, by comparing $E_b$ with the energy of a square bending into part of a cylinder with known curvature: $k_b = 1.03K_B$. The angle $\theta_{0}$ is zero as we assume a flat equilibrium state. We also denote the sum of deflections on the sheet as a wrinkling parameter $B = \sum_{\mathrm{adj}\ \alpha,\beta} \left(1-\cos(\theta_{\alpha\beta} - \theta_0)\right)$, which simply represents the extent of wrinkling on the sheet surface and is related to the total bending energy by the bending constant: $E_b = k_b B$.

For a sedimenting particle, the body force due to gravity acting on the sheet is described by a potential $U$. 
\begin{equation}
    U = 4\Delta \rho g a^2 h_0\Delta x.
\end{equation}
We also include a truncated Lennard-Jones(LJ) repulsive potential to prevent the sheet from self-intersecting. 
Numerically, the potential acts between all the nodes, except for the nodes that are located within 3-ring neighbors of the targeted node at the equilibrium state. The potential has the form:
\begin{equation}
	E_{LJ}= 
	\begin{cases}
		4 \varphi_0\left[\left(\frac{\sigma}{r}\right)^{12}-\left(\frac{\sigma}{r}\right)^{6}\right],	&\text{if } r<\sigma\\
		0,	&\text{otherwise}
	\end{cases}
\end{equation}
where $r$ is a instant distance between nodes, $\sigma$ scales the range of the potential, and $\varphi_0$ measures is the strength of the potential. Here, we choose $\sigma = 0.06a$ and $\varphi_0 = 4\times10^{-6}$. With these parameters, the only case where the repulsive force is active is when the sheet edge would otherwise cut through the surface, which only occurs for sheets with very small bending stiffness.
In addition, we also apply a similar truncated LJ potential to prevent sheets from entering a distance $d_\mathrm{lim}$ from the wall, and the potential only acts along the wall-normal direction in the range $d < d_\mathrm{lim}$. It is worth noting that, this interaction only acts when the sheet fully deposits on the wall and has no other effect on near-wall dynamics.

We introduce two nondimensional parameters to describe the aforementioned mechanical properties. The capillary number $\Ca = {\Delta \rho g a h_0}/{G}$ compares gravity and the in-plane elastic response of the sheet. Larger $\Ca$ indicates that the sheet is more deformable. In this study, we mainly focus on an almost inextensible sheet, such that the in-plane deformation is strongly constrained with a small capillary number $\Ca = 0.01$. We have verified that a further decrease in $\Ca$ (by 10 times) results in no qualitative change in sedimentation conformations. The comparison between bending and gravity is measured by elasto-gravitational number $\eg={\Delta \rho g a^3 h_0}/{K_B}$: larger $\eg$ indicates that the sheet is more flexible. In this study, we mainly focus on the influence of $\eg$.

\subsection{Numerical methods}
The numerical method for the elasticity problem is adapted from Charrier $\textit{et al.}$ \cite{charrier1989free}. We simulate the sheet by keeping track of nodes that move as material points on the sheet surface. 
From the elastic energies, we obtain the summed nodal elastic force ($\mathbf{F}_{e,i} = \mathbf{F}_{s,i}+\mathbf{F}_{b,i}$) exerted on each node from the first variation of the total energy with respect to the nodal displacements.
Each discretized element of the sheet is assumed to have homogenous deformation, so the element edges always remain linear. The deformed element is compared to its equilibrium shape under a local coordinate transformation via a rigid body rotation, and the displacement for any point inside the element is obtained by linear interpolation from the nodal positions. 
The detailed implementation can be found in references~\cite{yu2021coil,yu2022wrinkling,Fedosov2010,fedosov2010multiscale}. 
The total nodal force $\mathbf{F}_{e,i}$ is evaluated by summing the elastic force $(\mathbf{F}_{e,i})_j$ due to the deformation of each surrounding element $j$ shared by the node $i$: $\mathbf{F}_{e,i} = \sum_j (\mathbf{F}_{e,i})_j$, where the sum is over all elements meeting at the node.

Besides the elastic force, we also need to consider the external body force $\mathbf{F}_{g}$. The body force acting on the sheet has been assigned to each node based on the Voronoi area around the node at the equilibrium state. Due to the conservation of mass, the nodal body force $\mathbf{F}_{g,i}$ remains constant in the simulation acting along the $x$ direction, where $\hat{\mathbf{e}}_x$ is a unit directional vector:
\begin{equation}
    \mathbf{F}_{g,i}= A_{\mathrm{Voronoi},i}\Delta \rho h_0 g \hat{\mathbf{e}}_x.
\end{equation}

For the results shown, we discretize the square sheet with 1152 elements and 625 nodes. We have verified that changes in mesh resolution lead to only small quantitative changes in the results and no qualitative changes.

\subsection{Fluid solver}

In the present work, we consider a small, thin square sedimenting at low Reynolds number. Under these assumptions, the inertia of the fluid at the particle scale is negligible, and the fluid is governed by the Stokes equation. For each node on the sheet surface, the elastic force $\mathbf{F}_{e,i}$ exerted by the sheet on the fluid, the external body force $\mathbf{F}_{g,i}$ acting on the sheet are balanced with hydrodynamic force $\mathbf{F}_{h,i}$ from the fluid:
\begin{equation}
    \mathbf{F}_{e,i} + \mathbf{F}_{g,i} + \mathbf{F}_{h,i} =\mathbf{0}.
\end{equation} 
To account for the fact that the forces are not completely localized to the nodal positions, we use the method of regularized Stokeslets to solve for the fluid motion~\cite{cortez2005method}: the force $\mathbf{F}_i = -\mathbf{F}_{h,i}$ exerted by node $i$ on the fluid corresponds to a regularized force density $\mathbf{f}^\kappa_i = \mathbf{F}_i \delta_\kappa(\mathbf{x}-\mathbf{X}_i)$, where $\mathbf{X}_i$ is the position of node $i$ and $\delta_\kappa(\mathbf{x})$ is a regularized delta function with regularization parameter $\kappa$. Thus, the governing equations are the Stokes equation with regularized nodal forces and the continuity equation:
\begin{equation}
    \begin{array}{c}
        {-\nabla p+\eta \nabla^{2} \mathbf{v}+\sum_i\mathbf{F}_i \delta_\kappa(\mathbf{x}-\mathbf{X}_i)=\mathbf{0}} \\ 
        {\nabla \cdot \mathbf{v}=0}.
    \end{array}
\end{equation}

\subsubsection{Free-space sedimentation}
In an unbounded fluid, the velocity field generated due to a regularized point force $\mathbf{f}= \mathbf{F}\delta_\kappa(\mathbf{x}-\mathbf{X}_i)$ can be represented using a regularized Stokeslet $\mathbf{G}_\kappa$~\cite{graham2018microhydrodynamics,yu2021coil,yu2022wrinkling},:
\begin{equation}
    \mathbf{v}_\kappa(\mathbf{x}) = \mathbf{G}_\kappa(\mathbf{x}-\mathbf{X}_i)\cdot \mathbf{F}.
\end{equation}
As $1/\kappa\rightarrow 0$, $\mathbf{G}_\kappa$ reduces to the usual Stokeslet operator $\mathbf{G}(\mathbf{x}-\mathbf{X}_i)=1/8\pi\eta r(\mathbf{I}+(\mathbf{x}-\mathbf{X}_i)(\mathbf{x}-\mathbf{X}_i)/r^2)$, where $r=||\mathbf{x}-\mathbf{X}_i||$.
There are many ways to regularize a delta function $\delta_\kappa(\mathbf{x})$; we choose a regularization function for which the difference between $\mathbf{G}$ and $\mathbf{G}_\kappa$ decays exponentially as $\kappa r\rightarrow\infty$~\cite{graham2018microhydrodynamics}:

\begin{equation}
    \delta_\kappa(r) = \frac{\kappa^3}{\sqrt{\pi}^3}\exp(-\kappa^2r^2)\left[\frac52-\kappa^2r^2\right].
    \label{eq:delta1}
\end{equation}
With this choice, 
\begin{equation}
    \mathbf{G}_{\kappa}(\mathbf{x}-\mathbf{X}_i)=\frac{\operatorname{erf}(\kappa r)}{8 \pi \eta r}\left(\mathbf{I}+\frac{(\mathbf{x}-\mathbf{X}_i)(\mathbf{x}-\mathbf{X}_i)}{r^{2}}\right)+\frac{\kappa e^{-\kappa^{2} r^{2}}}{4 \pi^{3 / 2} \eta}\left(\mathbf{I}-\frac{(\mathbf{x}-\mathbf{X}_i)(\mathbf{x}-\mathbf{X}_i)}{r^{2}}\right).
\end{equation}
In the simulations, $\kappa$ must be chosen to scale with the minimum node-to-node distance $l_{\mathrm{min}}$. We take $\kappa l_{\mathrm{min}} = 2.1842$, which is obtained from a validation case of a disc in biaxial extensional flow discussed in \cite{yu2021coil}. We represent the total velocity at a point $\mathbf{x}$ as 
\begin{equation}
    \mathbf{v}(\mathbf{x}) =\mathbf{v}_\infty(\mathbf{x}) +\mathbf{v}_p(\mathbf{x})= \mathbf{v}_\infty(\mathbf{x}) +\sum_i\mathbf{G}_\kappa(\mathbf{x}-\mathbf{X}_i)\cdot \mathbf{F}_i.
    \label{eq:vfree}
\end{equation}

\subsubsection{Near-wall sedimentation}
The presence of an infinite wall complicates the problem as we require an extra boundary condition for the fluid: the velocity needs to vanish on the no-slip wall ($\mathbf{v}(y = 0) = \mathrm{0}$ assuming a vertical wall)~\cite{blake1971note}. Therefore, the velocity field generated to a regularized point force is now accounted for by the regularized Blakelet to satisfy this boundary condition~\cite{ainley2008method, cortez2015general}. The regularized Blakelet consists of a regularized Stokeslet and an image system across the wall which contains a combination of multipolar expansion terms from the Stokes equation that together cancel the induced velocity on the wall to enforce no-slip condition. Compared to singular solution derived by Blake, we need two extra rotlets to account for residual terms due to regularization. Following Ainley \textit{et al.}~\cite{ainley2008method}, we apply two types of regularized delta functions: $\delta_\kappa(\mathbf{x})$ and $\delta^d_\kappa(\mathbf{x})$ to regularize different terms in the image system. 
Here, $\delta_\kappa(r)$ is from Eq.~\ref{eq:delta1}, and $\delta^d_\kappa(r)$ has the form:
\begin{equation}
    \delta^d_\kappa(\mathbf{x}) = \frac{\kappa^3}{\sqrt{\pi}^3}\exp(-\kappa^2r^2).
\end{equation}
The two functions satisfy the relation:
\begin{equation}
    \delta_\kappa(\mathbf{x}) = \frac12\left(r\frac{\partial\delta^d_\kappa(\mathbf{x})}{\partial r}+5\delta^d_\kappa(\mathbf{x})\right).
\end{equation}
We define $\mathbf{r} = \mathbf{x}-\mathbf{X}_i$ and $\mathbf{R} = \mathbf{x}-\mathbf{X}^I_i$, where $\mathbf{X}_i$ is the the position of the node $i$ and $\mathbf{X}^I_i$ is the corresponding image point. Therefore, the regularized Blakelet $\mathbf{B}_{\kappa}$, in index notation, can be expressed as:
\begin{equation}
\begin{aligned}
B_{\kappa,i j}\left(\mathbf{r}\right)&= G_{\kappa,i j}\left(\mathbf{r}\right)-G_{\kappa,i j}\left(\mathbf{R}\right)\\
&-2 y_{0}\left(\delta_{j l}-2\delta_{j 2}\delta_{2 l}\right)
\left(D_{\kappa, i2l}(\mathbf{R})-\frac{y_0}{2\eta} P^d_{\kappa, il}(\mathbf{R})\right)\\
&+2 y_{0}\left(R_{\kappa,il}(\mathbf{R})-R^d_{\kappa,il}(\mathbf{R})\right)\epsilon_{jl2}.
\end{aligned}	
    \label{eq:vBlake}
\end{equation}
where the regularized Stokeslet $\mathbf{G}_{\kappa}\left(\mathbf{r}\right)$, the image regularized Stokeslet $\mathbf{G}_{\kappa}\left(\mathbf{R}\right)$, the regularized Stokes doublet $\mathbf{D}_{\kappa}\left(\mathbf{R}\right)$ and the rotlet $\mathbf{R}_{\kappa}\left(\mathbf{R}\right)$ are regularized by function $\delta_\kappa(\mathbf{x})$. The regularized potential dipole $\mathbf{P}_{\kappa}\left(\mathbf{R}\right)$ and the second rotlet $\mathbf{R}^d_{\kappa}\left(\mathbf{R}\right)$ are regularized by the alternative function $\delta^d_\kappa(\mathbf{x})$.
Note that if $1/\kappa\rightarrow 0$, the solution recovers the singular Blakelet as two rotlets cancel each other~\cite{ainley2008method, cortez2015general}. Here,

\begin{equation}
\begin{aligned}
G_{\kappa, i j}(\mathbf{r})
&= \frac{1}{8 \pi \eta} \left[\left( \dfrac{\operatorname{erf}\left(\kappa r\right)}{r}+\dfrac{2\kappa \mathrm{e}^{-\kappa ^2r}}{\sqrt{{\pi}}} \right)\delta_{i j} \right.
+\left.\left( \dfrac{\operatorname{erf}\left(\kappa r\right)}{r}-\dfrac{2\kappa \mathrm{e}^{-\kappa ^2r}}{\sqrt{{\pi}}} \right) \frac{r_i r_j}{r^{2}}\right],
\end{aligned}	
\end{equation}

\begin{equation}
\begin{aligned}
G_{\kappa, i j}(\mathbf{R})
&= \frac{1}{8 \pi \eta} \left[\left( \dfrac{\operatorname{erf}\left(\kappa R\right)}{R}+\dfrac{2\kappa \mathrm{e}^{-\kappa ^2R}}{\sqrt{{\pi}}} \right)\delta_{i j} \right.
+\left.\left( \dfrac{\operatorname{erf}\left(\kappa R\right)}{R}-\dfrac{2\kappa \mathrm{e}^{-\kappa ^2R}}{\sqrt{{\pi}}} \right) \frac{R_i R_j}{R^{2}}\right],
\end{aligned}	
\end{equation}

\begin{equation}
\begin{aligned}
D_{\kappa, i 2 l}(\mathbf{R})
&= \frac{1}{8 \pi \eta}
\left[\frac{1}{R^2}
\left( \dfrac{\operatorname{erf}\left(\kappa R\right)}{R}-\dfrac{2\kappa \mathrm{e}^{-\kappa ^2R^2}}{\sqrt{{\pi}}} \right)
\left(\delta_{2 l} R_{i}-\delta_{i 2}R_l + R_2\left(\delta_{i l}-3 \frac{R_{i} R_{l} }{R^{2}}\right)\right)
\right. \\
&\left.
-2\kappa^2\dfrac{2\kappa \mathrm{e}^{-\kappa ^2R^2}}{\sqrt{{\pi}}}\left(\delta_{i 2}R_l - R_2\frac{R_i  R_l}{R^{2}}\right)
\right],
\end{aligned}	
\end{equation}

\begin{equation}
\begin{aligned}
P^d_{\kappa, i l}(\mathbf{R})
&= -\frac{1}{4 \pi} 
\left[\frac{1}{R^2}\left( \frac{\operatorname{erf}\left(\kappa R\right)}{R}- \frac{2\kappa (2\kappa^2 R^2+1)\mathrm{e}^{-\kappa ^2R^2}}{\sqrt{\pi}} \right) \left(\delta_{il}- \dfrac{3R_iR_l}{R^2}\right)\right.\\
&\left. +\dfrac{4\kappa \mathrm{e}^{-k^2R^2}}{\sqrt{\pi}R^2}\dfrac{R_iR_l}{R^2}\right],
\end{aligned}	
\end{equation}

\begin{equation}
\begin{aligned}
R_{\kappa, i l} (\mathbf{R})
=\left(\dfrac{\operatorname{erf}\left(\kappa R\right)}{4{\pi}R^3}+\dfrac{\kappa \left(\kappa^2R^2-1\right)\mathrm{e}^{-\kappa^2R^2}}{2{\pi}^\frac{3}{2}R^2}	 \right)\epsilon_{ilm}R_m,
\end{aligned}	
\end{equation}

\begin{equation}
\begin{aligned}
R^d_{\kappa, i l} (\mathbf{R})
=\left(\dfrac{\operatorname{erf}\left(\kappa R\right)}{4{\pi}R^3}-\dfrac{\kappa \mathrm{e}^{-\kappa ^2R^2}}{2{\pi}^\frac{3}{2}R^2}	\right)	\epsilon_{ilm}R_m.
\end{aligned}	
\end{equation}

Therefore, the total velocity at a point $\mathbf{x}$ as
\begin{equation}
    \mathbf{v}(\mathbf{x}) =\mathbf{v}_\infty(\mathbf{x}) +\mathbf{v}_p(\mathbf{x})= \mathbf{v}_\infty(\mathbf{x}) +\sum_i\mathbf{B}_\kappa(\mathbf{x}-\mathbf{X}_i)\cdot \mathbf{F}_i.
    \label{eq:vbdd}
\end{equation}

At each time step for either the free space or the near wall case, after updating the force balance, we update the nodal positions by applying the condition that they move with the local fluid velocity in Eq.~\ref{eq:vfree} or Eq.~\ref{eq:vbdd}, thereby satisfying the no-slip and no-penetration conditions:
\begin{equation}
	\frac{d\mathbf{X}_i}{dt}=\mathbf{v}(\mathbf{X}_i).
\end{equation}

This equation is solved with a forward-Euler method. We have verified the solution with other numerical methods such as the fourth-order Runge-Kutta method, and find that the higher-order method shows negligible effects on the results based on the timestep used.
For numerical stability, the timestep applied follows $\Delta t = 0.1\Ca l_{\mathrm{min}}$, and we enforce an upper limit of $5 \times 10^{-4}$ strain unit per time step. In all simulations, we initially assign the sheet random perturbation with a small magnitude ($10^{-4}$) to trigger potential deformations.

\newpage
\section{RESULTS AND DISCUSSION} \label{sec:sed_results} 
    \subsection{Free space sedimentation} \label{sec:results_sed_free}


\begin{figure}[!t]
    \centering
    \captionsetup{justification=raggedright}
   \includegraphics[width=0.8\textwidth]{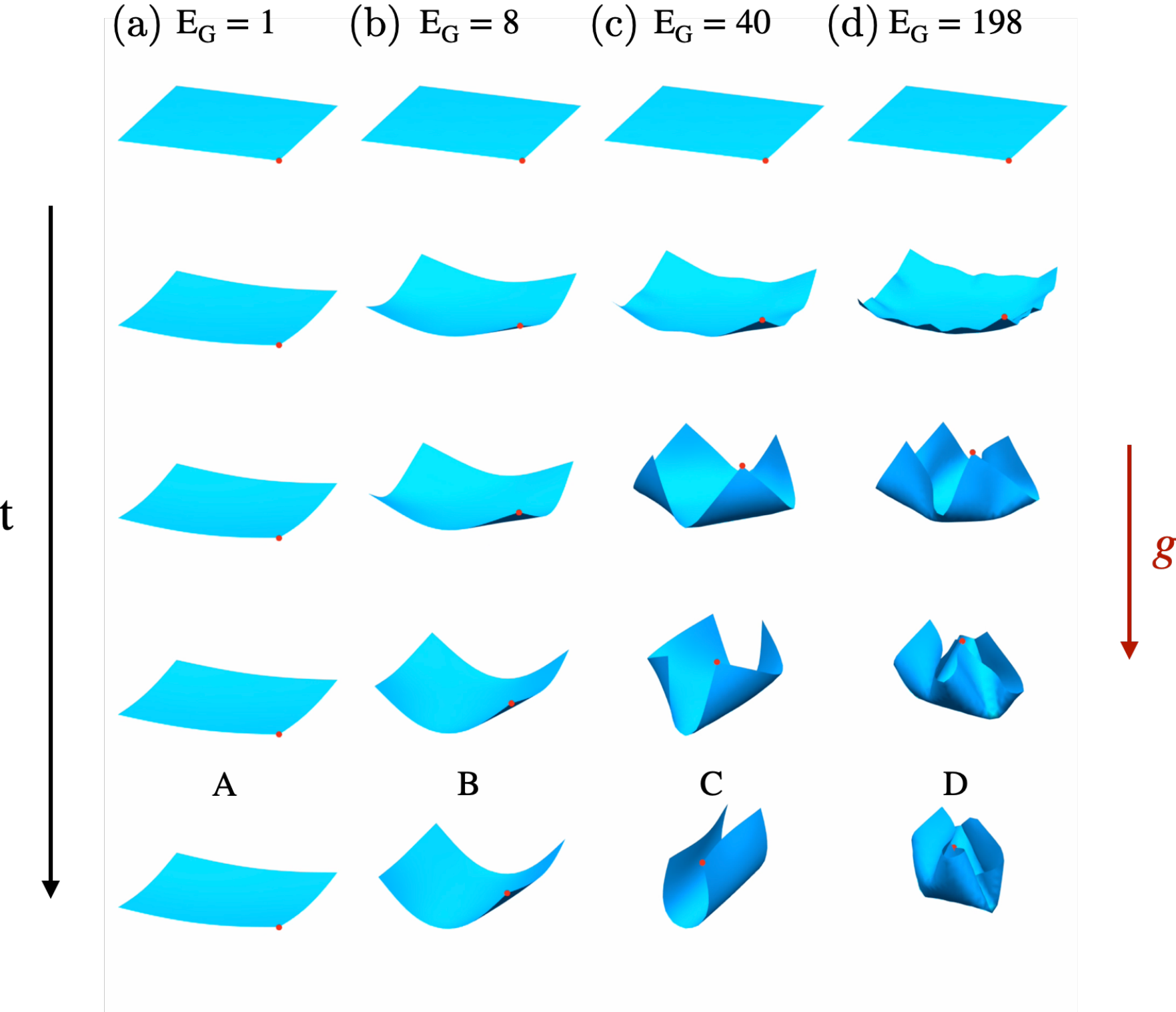}
    \caption{Snapshots of a square sheet sedimenting in free space with $\phi = 90^\circ$. (a) Flat ($\eg = 1$). (b) Convex taco ($\eg = 8$). (c) Concave taco ($\eg = 40$). (d) Complex ($\eg = 198$). A-D represent the final steady conformation. Snapshots only show the conformation evolution without actual sedimenting distance. See Supplemental Material at [URL will be inserted by publisher: Appendix \ref{sec:appendix} Movie 1] for animated movies.} 
    \label{fig:res_free_snap}    
\end{figure} 

We first consider a sheet sedimenting in free space. The initial orientation of the sheet aligns perpendicular to gravity ($\phi = 90^\circ$). In Fig.~\ref{fig:res_free_snap}, we show four distinct examples with increasing $\eg$ to address the effect of sheet flexibility. 
For the stiff sheet shown in Fig.~\ref{fig:res_free_snap}(a), sheet stiffness dominates over gravity. The sheet settles with an almost flat orientation, with negligible deformation.
With increasing $\eg$, sheet deformability becomes more pronounced. Due to hydrodynamic interactions between different parts of the sheet surface, the center part of the sheet initially sediments fastest, which drives the four corners to symmetrically lift upward. This metastable state soon becomes unstable and two edges fold inward while the other two edges relax to form a taco shape.
Based on $\eg$, we observed two types of tacos. A less flexible sheet (Fig.~\ref{fig:res_free_snap}(b)) indicates a convex taco. The slightly folded edges form a V shape, and the relaxed edges are convex. A more flexible sheet (Fig.~\ref{fig:res_free_snap}(c)) ends up with a concave taco. The folded edges become a U shape, and the relaxed edges are concave. 
For very flexible sheets as Fig.~\ref{fig:res_free_snap}(d), very small resistance to bending leads to a complex crumpled geometry that is not considered in this study.

\begin{figure}[t]
    \centering
    \captionsetup{justification=raggedright}
\includegraphics[width=\textwidth]{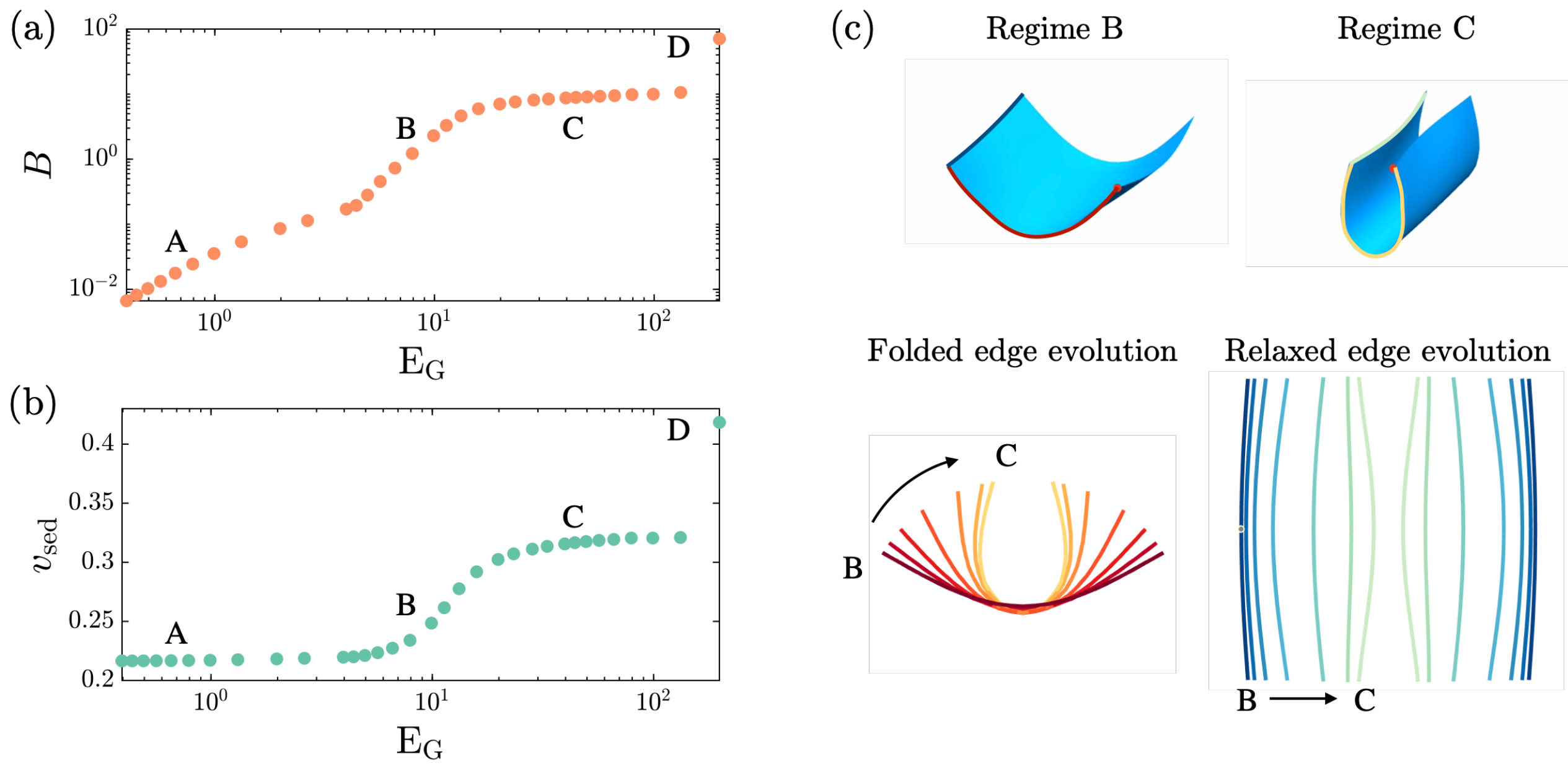}    \caption{Characterization of sedimentation dynamics. A-D are steady conformations shown in Fig.~\ref{fig:res_free_snap} representing four distinct regimes. (a) Evolution of wrinkling parameter $B$ on the sheet vs. $\eg$. (b) Terminal velocity $v_\mathrm{{sed}}$ vs. $\eg$. (c) The transition of conformations between regime B and regime C is shown as the evolution of folded and relaxed edges.}
    \label{fig:res_free_char}    
\end{figure} 

We characterize the sedimentation dynamics by examining the folding of the sheet and the sedimentation velocity of the final steady conformations. In Fig.~\ref{fig:res_free_char}(a), we show the extent of folding by calculating the wrinkling parameter $B$ defined earlier (the sum of deflection between all neighboring elements) for sheets over a range of $\eg$. 
In Fig.~\ref{fig:res_free_char}(b), we quantify the sedimentation dynamics by measuring a terminal velocity $v_\mathrm{sed}$ as shown in Fig.~\ref{fig:res_free_char}(b). 
In both plots, we observe four different regimes, corresponding to the four examples (A-D) given in Fig.~\ref{fig:res_free_snap}.  
Typically, the transitions of shape conformations resemble flexible filament sedimentation with increasing flexibility~\cite{cunha2022settling,li2013sedimentation}. Sheets in regime A show negligible deformation and maintain an almost flat conformation, which leads to larger drag and slower sedimentation. With increasing $\eg$, the evolution of relaxed edges in Fig.~\ref{fig:res_free_char}(c) demonstrates a smooth transition from a relatively unfolded state (regime B) to a compact folded state (regime C). Elastic sheet conformations are more complex than fiber. With the transition of folded edges, sheets also show a transition of the relaxed edges from convex to concave, in order to accommodate a less wrinkled state. Finally, regime D represents a fully wrinkled shape and compact conformation sediments faster in free space.
In summary, sheets with increasing $\eg$ show more evident folding behavior and the conformations become more compact. The more compact conformations experience smaller drag and sediment faster.


\begin{figure}[t]
    \centering
    \captionsetup{justification=raggedright}
    \includegraphics[width=\textwidth]{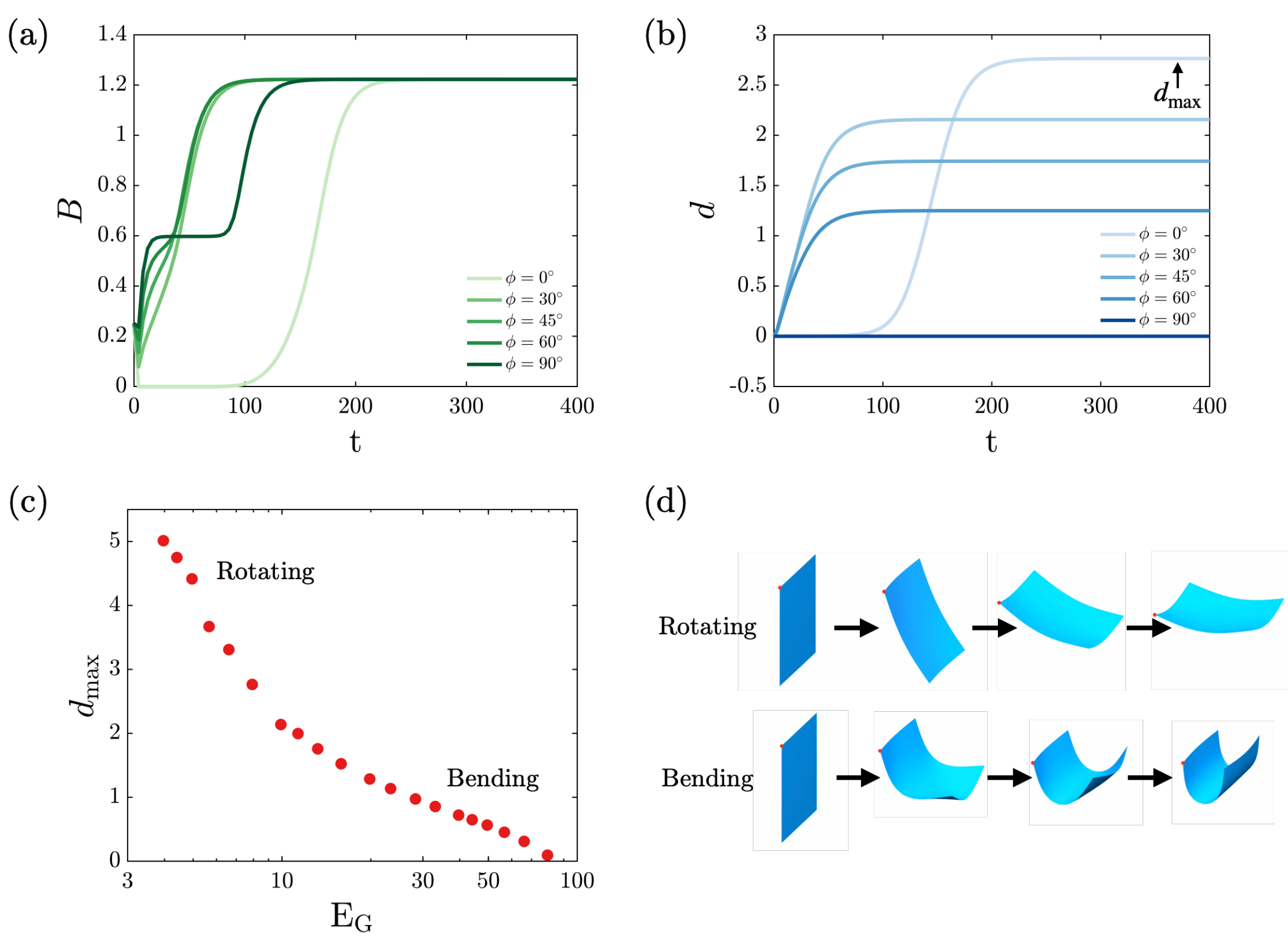}
    \caption{Influence of initial orientation $\phi$ and $\eg$ on sheet reorientation dynamics: (a) Transient evolution of wrinkling on the sheet for $\eg = 8$ with different initial orientation $\phi$. (b) Transient evolution of lateral drift distance $d$ on the sheet for $\eg = 8$ with different initial orientation $\phi$. The maximum drift distance is denoted $d_\mathrm{max}$. (c) Maximum lateral distance $d_\mathrm{max}$ for sheets with different $\eg$. (d) Illustration of two types of reorientation dynamics: rotating ($\eg = 4$) and bending ($\eg = 13$) from selected cases in Fig.~\ref{fig:res_free_reorient}(c). The sheet is initially oriented at $\phi = 0 ^\circ$ subject to small and random perturbations. See Supplemental Material at [URL will be inserted by publisher: Appendix \ref{sec:appendix} Movie 2] for animated movies.}
    \label{fig:res_free_reorient}    
\end{figure} 

\add{The conformations in Fig.~\ref{fig:res_free_snap} represent the final shapes for flexible sheets with random initial orientations (i.e. the conformation is independent of $\phi$.). The only exception is when a flat sheet is initially aligned perfectly with gravity, in which case it reaches a flat state that is unstable to small perturbations.}
Here, we illustrate the independence of initial orientation $\phi$ on the final conformation by examining case B of Fig.~\ref{fig:res_free_snap} ($\eg = 8$). 
Figure \ref{fig:res_free_reorient}(a) shows the evolution of the wrinkling parameter $B$ for various initial orientation angles, indicating that all $\phi$ reach the same final conformation shown in Fig.~\ref{fig:res_free_snap}. Though arriving at the same conformation, different initial conditions exhibit significantly different transient reorientation dynamics, shown via the evolution of lateral drift distance $d$ in Fig.~\ref{fig:res_free_reorient}(b). 
There is no drift observed in the case $\phi = 90^\circ$, which implies a stable orientation. The drift distance $d$ increases with decreasing of $\phi$ and the maximum drift happens when the sheet is initially parallel to the gravity ($\phi = 0^\circ$). Therefore, by the symmetry of free space, all possible sedimenting trajectories of the sheet are restricted to a cloud. This observation agrees with observations in flexible filaments~\cite{li2013sedimentation}. We denote the maximum lateral drift distance (which occurs when $\phi = 0^\circ$) as $d_\mathrm{max}$. 

We summarize $d_\mathrm{max}$ for sheets with different $\eg$ in Fig.~\ref{fig:res_free_reorient}(c). The drift due to reorientation is strongly influenced by $\eg$, as more flexible sheets show smaller $d_\mathrm{max}$. 
After carefully examining the dynamics, we introduce two categories of different reorientation dynamics, which we denote rotating and bending, with examples illustrated in Fig.~\ref{fig:res_free_reorient}(d). Here, all sheets initially stay parallel to gravity subject to small random perturbations.
For stiff sheets with $\eg \lesssim 6$, perturbation induces small deformation on the sheet while the sheet stays almost flat. However, the small deformability induces the orientation of the sheet to slowly align perpendicular to gravity. During reorientation, the sheet rotates around the sheet center while the asymmetric and tilted conformation induces drift along the lateral direction resulting in evident translation. 

Unlike the stiff sheets, the flexible sheets with $\eg \gtrsim 6$ bend during reorientation. Here, the upper and lower part of the sheet shows relatively independent deformation as illustrated in Fig.~\ref{fig:res_free_reorient}(d); the upper part bends slightly, and the lower part of the sheet folds and rotates relatively to the sheet center to form a taco shape. Bending sheets exhibit smaller lateral drift compared to rotating ones. For larger $\eg$, the flexibility ensures a quick folding reorientation and the drift becomes negligible.
We summarized the evolution of maximum drift distance $d_\mathrm{max}$ with $\eg$ in Fig.~\ref{fig:res_free_reorient}(c), where the two different reorientation types can be recognized. 
Similar orientation dynamics were also reported in reorientation dynamics of flexible filaments~\cite{lagomarsino2005hydrodynamic,li2013sedimentation}.



\subsection{Sedimentation next to a vertical wall} \label{sec:results_sed_vwall}

In this section, we place the sheet next to an infinite wall, which can be adjusted with an angle $\alpha$. 
We first illustrate the dynamics next to a vertical wall ($\alpha = 0 ^\circ$) and apply the initial orientation $\phi = 90 ^\circ$, which is the stable orientation in free space.
The sheet is initially a distance $d_0 = 1.5$ away from the wall. 

\begin{figure}[t]
    \centering
    \captionsetup{justification=raggedright}
    \includegraphics[width=\textwidth]{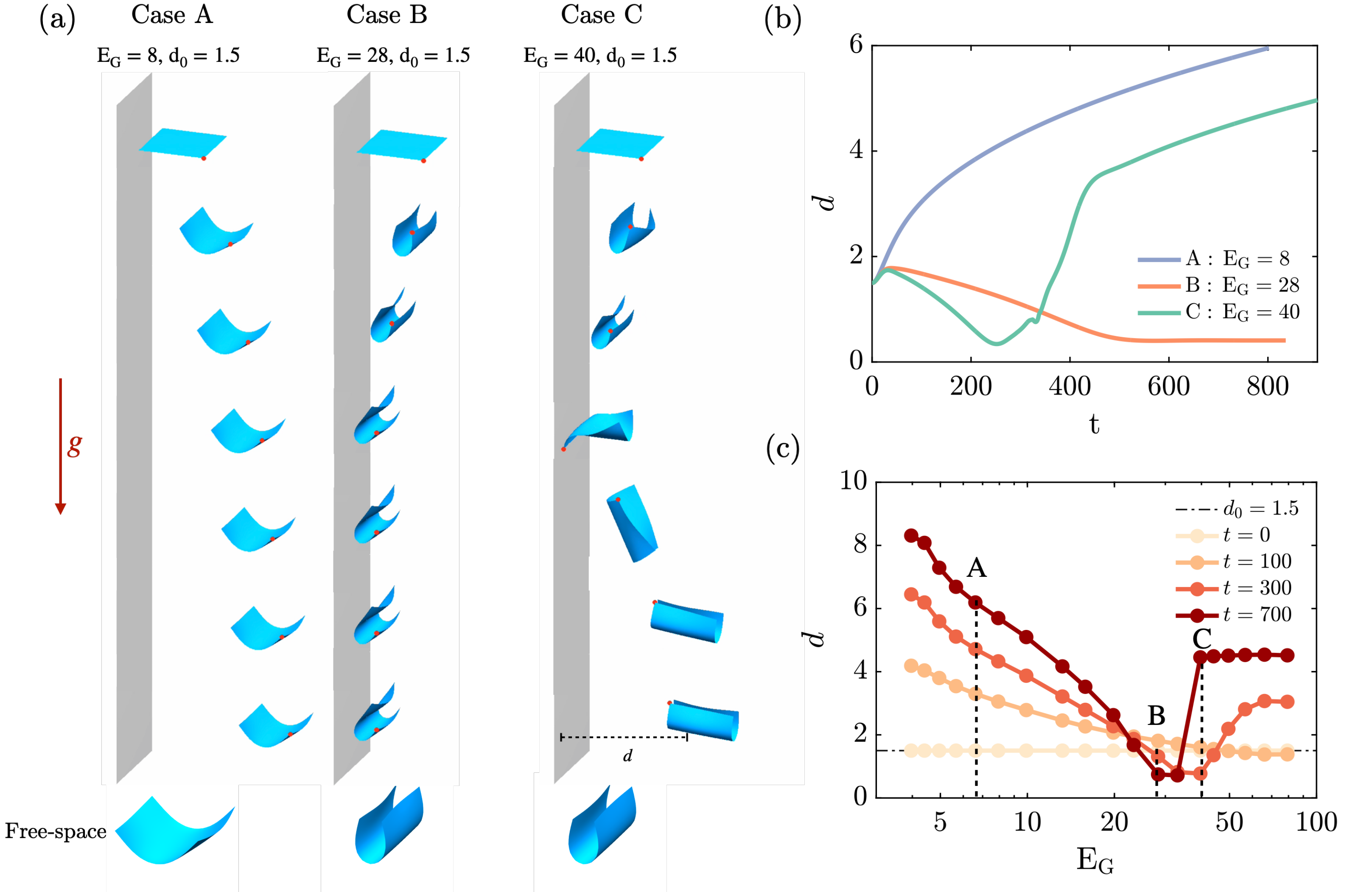}
    \caption{(a) Snapshots of dynamics near a vertical wall with initial orientation $\phi = 90^\circ$ and $d_0 = 1.5$. Case A: $\eg = 8$. Case B: $\eg = 28$. Case C: $\eg = 40$. The transient migration distance away from the wall is denoted $d$. Each case has its steady free space conformation for comparison. See Supplemental Material at [URL will be inserted by publisher: Appendix \ref{sec:appendix} Movie 3] for animated movies. (b) Transient evolution of migration distance $d$ for three sheets in Fig.~\ref{fig:res_wall_vertical}(a). (c) Temporal evolution of migration distance $d$ for sheets with different $\eg$, with labels indicating cases shown in Fig.~\ref{fig:res_wall_vertical}(a). See Supplemental Material at [URL will be inserted by publisher: Appendix \ref{sec:appendix} Movie 4] for animated movies. } 
    \label{fig:res_wall_vertical}    
\end{figure} 

The presence of the wall breaks spatial symmetry, resulting in an asymmetric sheet conformation in contrast to a symmetric conformation in free space. 
In Fig.~\ref{fig:res_wall_vertical}(a), we show the dynamics of three cases with increasing $\eg$, where free space evolutions of the sheets are symmetric tacos. 
The asymmetric sheet, due to anisotropic drag, exhibits drift of the sheet along the wall-normal (lateral) direction while sedimenting. This leads to the migration of the sheet away from the wall where the migration dynamics are strongly determined by $\eg$. 

In the short term, the relatively stiff sheet (case A $\eg = 8$ in Fig.~\ref{fig:res_wall_vertical}(a)) directly migrates away from the wall. In contrast, the other two flexible sheets (case B and case C) initially migrate toward the wall. The noticeable difference in dynamics arises from the sheet conformations. In case A, small $\eg$ leads to a slightly folded sheet where elasticity dominates and the overall orientation of the asymmetric taco tilts away from the wall due to the unbalanced side edges. The resultant shape induces the sheet to drift away from the boundary. In cases B and C, the sheets are more flexible and the conformations are more compact, with two side edges getting close to each other. 
Here, the folded side closer to the wall experiences stronger drag than the side away from the wall, leading to a tilted conformation that drifts toward the wall as it sediments.  

In the long term, the stiff sheet (case A in Fig.~\ref{fig:res_wall_vertical}(a)) continues migrating away, while flexible sheets (case B and case C in Fig.~\ref{fig:res_wall_vertical}(a)) reorient near the wall and take on a shape that eases migration. That is, the stronger wall-induced asymmetry near the wall triggers the outer side edge of the flexible sheet to turn inside out with a flip. Here, a sufficiently flexible sheet (case C) forms a taco with an overall orientation now tilting away from the wall, and the new orientation causes the sheet to drift away, as indicated in the snapshots of case C in Fig.~\ref{fig:res_wall_vertical}(a). Interestingly, we notice a small regime of intermediate $\eg$ (case B), where the sheet becomes trapped near the wall. More specifically, the folded sheet sediments next to the wall with an almost constant wall-normal distance, while the conformation shows weak oscillations as the conformation manages to reorient. 
The oscillation in conformation is very small in the snapshots of case B shown in Fig.~\ref{fig:res_wall_vertical}(a). Overall, the sheet is ``sliding'' next to the wall at a nearly fixed distance. It is worth noting that the sliding sheet is a result of both initial orientation and $\eg$. If the sheet is initially oriented away from the wall, it immediately triggers the migration and the sheet escapes away from the wall.

The above examples illustrate three distinct responses of a sheet next to a vertical wall, determined by $\eg$. To characterize the migration, we show in Fig.~\ref{fig:res_wall_vertical}(b) the evolution of wall-normal distance $d$. The flexible sheet (case C) migrates slower compared to the stiffer sheet (case A) as it first drifts toward the wall and reorients to migrate away, indicating a non-monotonic increase in the wall-normal distance.  The intermediate case B becomes trapped near the wall. 

To elaborate on the migration dynamics, we examine the evolution of $d$ in the parameter space of $\eg$ in Fig.~\ref{fig:res_wall_vertical}(c). At short times ($t<100$), migration distance decreases with $\eg$ and very flexible sheets begin to drift toward the wall.
When $t = 300$, flexible sheets reorient near the wall and migrate away, while sheets with intermediate $\eg$ become trapped around the wall, resulting in a non-monotonic trend in $d$. 
Here, we identified two transition regimes of $\eg$: $\eg \sim 23$ shows the threshold when the sheet immediately migrates away from the wall, and $23 < \eg < 28$ marks the onset where the sheet can successfully reorient and escape. 
In addition, we examined the influence of initial distance by placing the sheet away from the wall with $d_0 = 3$ and the results are in qualitative agreement with the near wall condition. Since the wall effect is long-ranged, flexible sheets initially drift toward the wall such as case B and C in Fig.~\ref{fig:res_wall_vertical}(a) follows the same dynamics even when the sheet is initially further away. 

 Finally, we consider the migration with a simple analysis. 
 When considering the multipole expansion of a particle sedimenting near a vertical wall, the total force is vertical (downward). Here, the Stokeslet contribution to the particle velocity coming from the Blakelet is also vertical (but upward), and leads to no wall-normal migration. Thus, the leading order contribution of the Stokeslet image to the particle migration must be dipolar, scaling as  
 $1/d^2$. Therefore, the migration velocity along the wall-normal direction $\mathrm{v}=\mathrm{d} d/\mathrm{d}t\sim 1/d^2$ when the sheet is relatively away from the wall. Integrating the equation with the initial condition $d(t = 0) = d_0$, we obtain a simple relation between the wall-normal distance $d$, time $t$, and a fitted constant $k$:
\begin{equation}
    d = (3kt+{d_0}^3)^{1/3}.
    \label{eq: scailing}
\end{equation}
To test the scaling, we take case A in Fig.~\ref{fig:res_wall_vertical}(a) and use the evolution of $d$ to fit Eq.~\ref{eq: scailing}.
\begin{figure}[t]
    \centering
    \captionsetup{justification=raggedright}
    \includegraphics[width=0.7\textwidth]{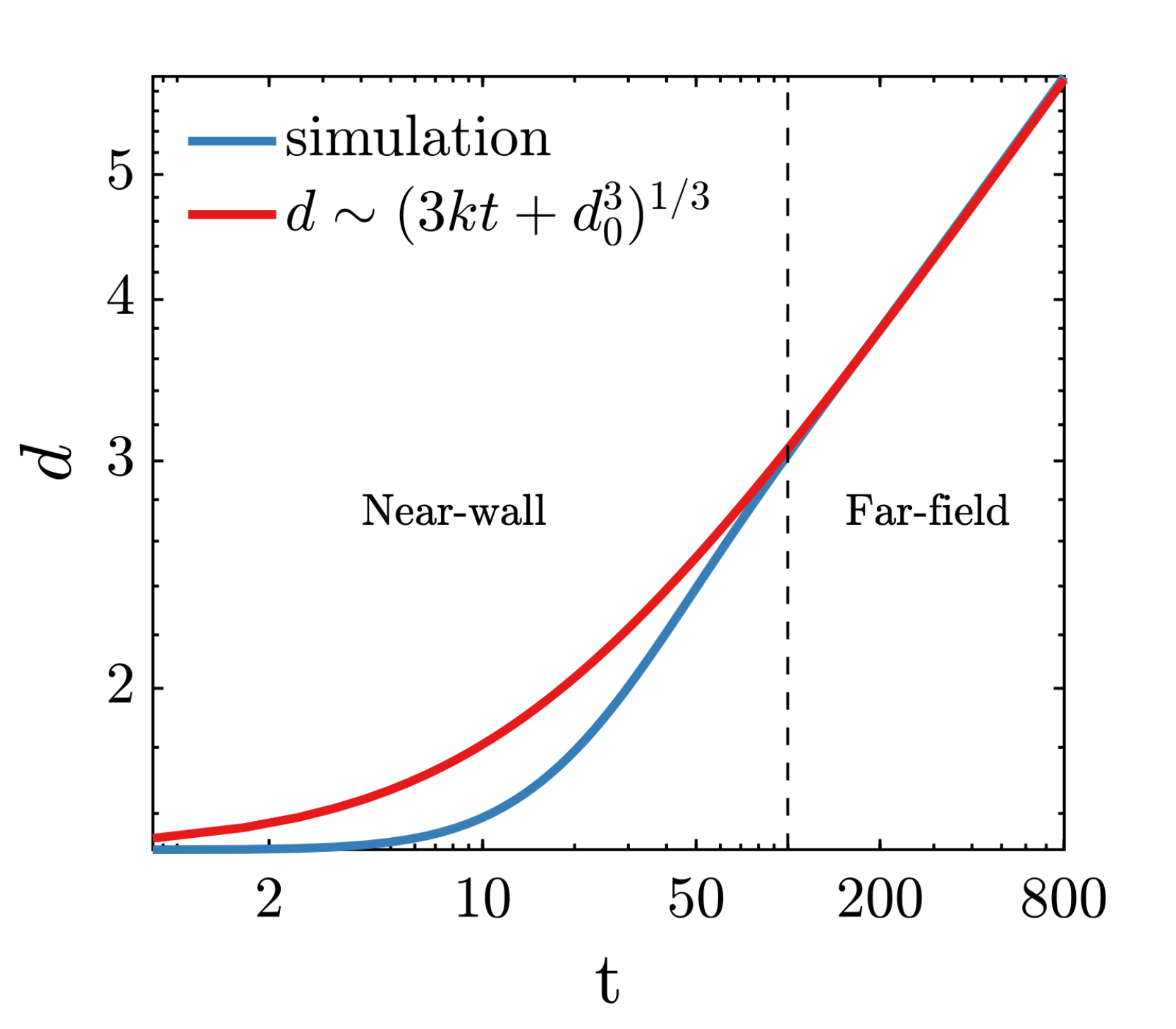}
    \caption{Comparison of wall-normal distance evolution for case A ($\eg = 8$) in Fig.~\ref{fig:res_wall_vertical}(a) by a fitting equation $d = (3kt+{d_0}^3)^{1/3}$ with $k = 0.085$ and $d_0 = 1.5$. We consider the dynamics for $t<100$ to be governed by near-wall effects.} 
    \label{fig:res_wall_vertical_scailing}    
\end{figure} 
The comparison in Fig.~\ref{fig:res_wall_vertical_scailing} shows that the simple scaling gives a good power-law fit when the sheet is further away from the wall ($t > 100$). The deviation from the fitting curve for $t < 100$ has two contributions. The first is near-wall hydrodynamics, and the other is the change in conformation as the sheet deforms from a flat state to adopt a taco conformation.


\newpage
\subsection{Sedimentation next to a tilted wall} \label{sec:results_sed_twall}

\begin{figure}[!htp]
    \centering
    \captionsetup{justification=raggedright}
    \includegraphics[width=\textwidth]{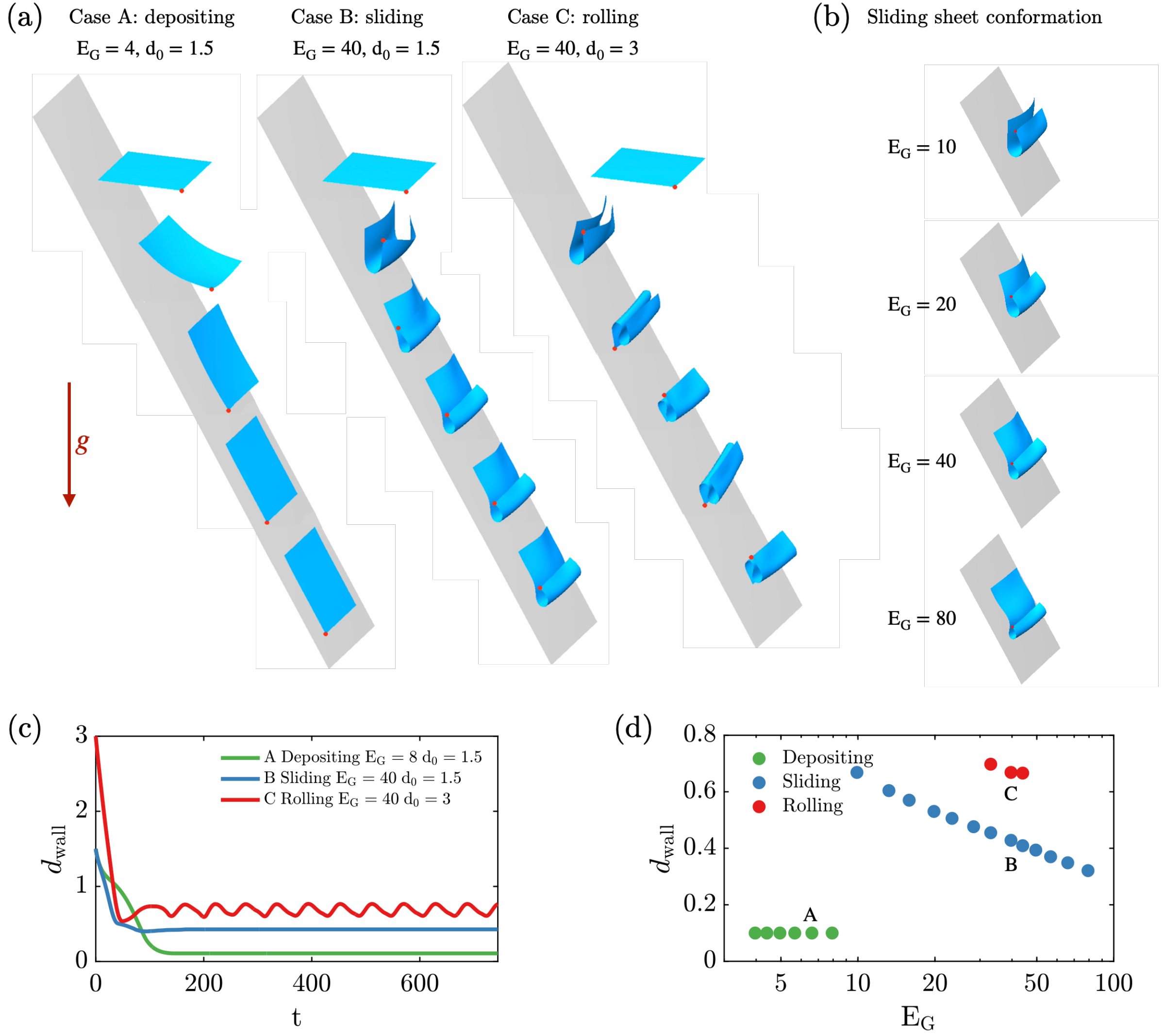}
    \caption{(a) Snapshots of dynamics near a tilted wall ($\alpha = 30^\circ$) with initial orientation $\phi = 90^\circ$: Case A. $\eg = 1$: the sheet deposits on the wall. Case B. $\eg = 40$: the sheet initially close to the wall ($d_0 = 1.5$) reaches a steady conformation and slides along the wall. Case C. $\eg = 40$: the sheet initially far from the wall ($d_0 = 3$) folds and rolls along the wall. See Supplemental Material at [URL will be inserted by publisher: Appendix \ref{sec:appendix} Movie 5] for animated movies. (b) Shape evolution of steady sliding sheets with increasing $\eg$. (c) Transient evolution of wall-normal distance $d$ for cases shown in Fig.~\ref{fig:res_tilted_a30snap}(a). (d) Steady wall-normal distance $d_{s}$ for sheets next to a tilted wall with $\alpha = 30^\circ $.} 
    \label{fig:res_tilted_a30snap}    
\end{figure}

Compared to the vertical wall, the tilted wall with $\alpha > 0^\circ$ introduces more complex sedimenting dynamics. 
We initially set $\alpha=30 
 ^\circ$ and illustrate the effects of $\eg$ on sedimentation dynamics in Fig.~\ref{fig:res_tilted_a30snap}. In summary, we find three types of dynamics, depending on $\eg$, which we denote ``depositing" (case A), ``sliding" (case B), and ``rolling" (case C). 
When the wall is tilted, gravity pulls the sheet toward the wall. Relatively stiff sheets ($\eg = 1$ Case A in Fig.~\ref{fig:res_tilted_a30snap}(a)) simply deposit on the wall.
In simulations, we apply a repulsive potential to prevent direct contact between the sheet and wall surface, such that the relaxed sheet becomes flat and sediments with a small fixed distance matching with the repulsive potential range. This is indicated in the evolution of the wall-normal distance $d$ in Fig.~\ref{fig:res_tilted_a30snap}(c) for case A.

For intermediate $\eg$, flexible sheets remain folded as they approach the tilted wall. There are two competing effects. The wall-induced asymmetric conformation drives the sheet to migrate away from the boundary as in the vertical wall case. Meanwhile, gravity pulls the sheet toward the wall. When these two effects are balanced, sufficiently flexible sheets may reach a steady conformation while sedimenting at a fixed distance away from the tilted wall, which we call sliding. The evolution of wall-normal distance reaches a fixed distance in Fig.~\ref{fig:res_tilted_a30snap}(c). 
We want to address the difference between sliding and depositing. Depositing is a result of added repulsive potential as we do not account for the contact between wall and sheets. In reality, the sheet will deposit on the wall without sedimenting. The sliding dynamics, however, result from the interplay between near-wall hydrodynamics, gravity, sheet elasticity, and wall angle. Removal of the repulsive potential does not affect sliding at all. 
The observation is analogous to the sliding dynamics reported for a rigid spheroid particle near a tilted wall observed by Mitchell \textit{et al.}~\cite{mitchell2015sedimentation}, where the spheroid sediments next to a tilted wall with a fixed orientation angle and keeps a fixed wall-normal distance. 
We illustrate several sliding conformations with increasing $\eg$ in Fig.~\ref{fig:res_tilted_a30snap}(b). To accommodate different hydrodynamic stress, the equilibrium sliding sheets adopt different conformations and wall-normal distances based on $\eg$. In general, stiffer sheets show less asymmetric conformation and stay a relatively large distance away from the wall. Flexible sheets show an inverse ``J'' shape, with folding mainly located around the lower part of the sheet and staying relatively close to the wall. 

In certain parameter regimes, sliding dynamics may be influenced by the initial distance from the wall. When $\alpha = 30^\circ$, we identified a small regime of $\eg$ where sheets show an interesting rolling motion when initially far from the wall. Case C in Fig.~\ref{fig:res_tilted_a30snap}(a) with $\eg = 40$ reveals an interesting rolling motion. The large wall initial distance ($d_0 = 3$) allows the sheet to deform while approaching the wall and fully fold into a compact S shape near the wall. The folded sheet then tumbles next to the wall resulting in an oscillatory trajectory as shown in Fig.~\ref{fig:res_tilted_a30snap}(c). In contrast, a near-wall initial condition ($d_0 = 1.5$) lacks space to fully fold on itself and ends up sliding, which indicates the sliding here is a result of hindered transient reorientation dynamics.

We summarize the different dynamics observed for $\alpha = 30^\circ$ by plotting the steady wall-normal distance $d_s$ for the equilibrium dynamics in Fig.~\ref{fig:res_tilted_a30snap}(d). When $\eg < 8$, sheets deposit on the wall as maintaining a small distance within the repulsive potential range. Sheets with $\eg > 8$ reach a steady conformation and slide along the wall, while the steady wall-normal distance $d_s$ decreases with $\eg$. Rolling only occurs in a small parameter regime when the sheet is initially away from the wall.

\begin{figure}[t]
    \centering
    \captionsetup{justification=raggedright}
    \includegraphics[width=\textwidth]{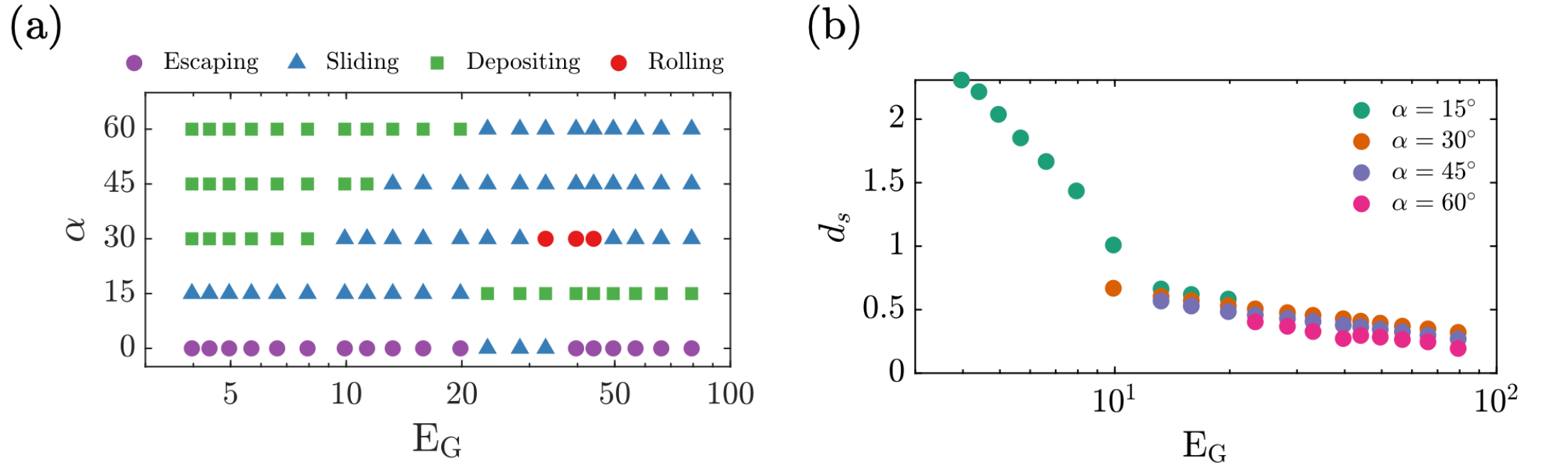}
    \caption{(a) Phase diagram for $\eg$ vs. $\alpha$. Symbols represent the sheet dynamics. (b)Steady wall-normal distance $d_s$ for sliding sheets (boxed region in Fig.~\ref{fig:res_wall_phase}(a)) with different $\eg$ and $\alpha$. } 
    \label{fig:res_wall_phase}    
\end{figure} 

To systematically illustrate the influence of flexibility $\eg$ and wall angle $\alpha$, a phase diagram of long-term dynamics is shown in Fig.~\ref{fig:res_wall_phase}(a). For every case, we tested two different $d_0$ to rule out dynamics due to the transient rolling state. 
Typically, when the wall is vertical ($\alpha = 0^\circ$), sheets escape the wall except for a small parameter regime (the sliding cases for $\alpha = 0^\circ$ in Fig.~\ref{fig:res_wall_phase}(a) correspond to case B in Fig.~\ref{fig:res_wall_vertical}(a)) where sheets are trapped due to reorientation. \add{When the wall is tilted, a very small $\alpha$ gives qualitatively similar dynamics as the vertical case. With a relatively large $\alpha$, sheets either become trapped near the wall region or deposited on the boundary. The specific transition $\alpha$ is not determined.} For $\alpha \ge 15 ^\circ$, we determined a transition $\eg$ to distinguish between sheets deposited on the wall and sheets trapped near the wall (marked as the boundary between sliding and depositing). One observation is that the threshold $\eg$ to deposit increases with $\alpha$. 
\add{Interestingly, under a small wall angle ($\alpha = 15^\circ$), the sheet deposits at large $\eg$. Here, very flexible sheets initially formed a compact shape may unravel and become flat when next to a slightly tilted wall.} 
In addition, in Fig.~\ref{fig:res_wall_phase}(b) we plot the steady wall-normal distance $d_s$ for the sliding sheets for different $\eg$ and $\alpha$. For all $\alpha$, $d_s$ decreases with $\eg$. 
The phase diagram provides a potential strategy to separate sheet-like particles. For example, to separate particles with different $\eg$, we may choose a wall angle such that the transition $\eg$ of specific $\alpha$ lies in between $\eg$ of particles. Consequently, stiffer particles deposit on the wall while more flexible particles slide and can be collected downstream.



\section{CONCLUSION} \label{sec:sed_conclusion}
In this work, we perform numerical simulations to systematically explore the sedimentation of an almost inextensible but flexible square sheet in a viscous fluid. An elasto-gravitational number $\eg$ is used to measure the ratio between gravity and bending elasticity. We addressed two scenarios: sedimenting in free space or beside an infinite planar wall, where the wall can be tilted. 

In free space, the sheet takes on a stable orientation perpendicular to gravity, with a unique steady conformation that is independent of initial orientations. Sheets adopt more compact shapes with increasing $\eg$, varying from almost flat to taco-like buckled shapes, and more folded shapes sediment faster as they experience smaller hydrodynamic drag. The randomly oriented sheets reorient toward the stable orientation, which induces lateral drift. The maximum drift happens when the sheet is initially aligned with gravity. We identified two different reorientation dynamics, and the drift distance due to reorientation increases with $\eg$. 

The presence of a wall introduces asymmetry to the sheet conformation, and the sedimentation dynamics are determined by both $\eg$ and the wall-tilt angle $\alpha$. When next to a vertical wall, asymmetry in sheet conformation leads to the migration of the sheet away from the wall. We found three different drifting dynamics determined by $\eg$, and observed a non-monotonic trend of migration distance evolution due to the different responses in complex buckled shapes. When the wall is tilted, wall-induced migration competes with gravity-induced deformation, leading to the observation of three types of dynamics: depositing, sliding, and rolling. We developed a phase diagram to show the dependence of dynamics on $\alpha$ and $\eg$, and determined the transition $\eg$ between depositing and sliding for different $\alpha$.

This study deepens our understanding of the sedimentation dynamics of sheet-like particles and provides several strategies for particle separation. Based on $\eg$, we may separate particles via different settling velocities, or different migration distances if next to a vertical wall. Typically, a more robust separation method is to adjust the wall tilted angle. Since we know the transition $\eg$, an appropriate wall angle can make the relatively stiff sheets deposit on the wall while the more flexible particles keep sliding.

\section*{Conflicts of interest}
There are no conflicts to declare.

\begin{acknowledgments}
This work was supported by the National Science Foundation under grant No.~CBET-1604767 and Vannevar Bush Faculty Fellowship ONR No. N00014-18-1-2865.
\end{acknowledgments}

\begin{appendix}
\appendix 
\section{Movies} \label{sec:appendix}
\begin{itemize}
	\item Movie 1: Evolution of four initially flat sheets sedimenting perpendicular to gravity ($\phi = 90^\circ$) with increasing $\eg$.
	\item Movie 2: Evolution of two reorientation dynamics: rotating ($\eg = 4$) and bending ($\eg = 13$) from selected cases in Fig.~\ref{fig:res_free_reorient}(c).	
	\item Movie 3: Evolution of three initially flat sheets sedimenting perpendicular to gravity ($\phi = 90^\circ$) next to a vertical wall with increasing $\eg$.
	\item Movie 4: Evolution of migration distance $d$ for sheets with increasing $\eg$.
        \item Movie 5: Evolution of three initially flat sheets sedimenting perpendicular to gravity ($\phi = 90^\circ$) next to a tilted wall ($\alpha = 30^\circ$) with increasing $\eg$.
\end{itemize}

\end{appendix}



\begin{thebibliography}{45}%
\makeatletter
\providecommand \@ifxundefined [1]{%
 \@ifx{#1\undefined}
}%
\providecommand \@ifnum [1]{%
 \ifnum #1\expandafter \@firstoftwo
 \else \expandafter \@secondoftwo
 \fi
}%
\providecommand \@ifx [1]{%
 \ifx #1\expandafter \@firstoftwo
 \else \expandafter \@secondoftwo
 \fi
}%
\providecommand \natexlab [1]{#1}%
\providecommand \enquote  [1]{``#1''}%
\providecommand \bibnamefont  [1]{#1}%
\providecommand \bibfnamefont [1]{#1}%
\providecommand \citenamefont [1]{#1}%
\providecommand \href@noop [0]{\@secondoftwo}%
\providecommand \href [0]{\begingroup \@sanitize@url \@href}%
\providecommand \@href[1]{\@@startlink{#1}\@@href}%
\providecommand \@@href[1]{\endgroup#1\@@endlink}%
\providecommand \@sanitize@url [0]{\catcode `\\12\catcode `\$12\catcode
  `\&12\catcode `\#12\catcode `\^12\catcode `\_12\catcode `\%12\relax}%
\providecommand \@@startlink[1]{}%
\providecommand \@@endlink[0]{}%
\providecommand \url  [0]{\begingroup\@sanitize@url \@url }%
\providecommand \@url [1]{\endgroup\@href {#1}{\urlprefix }}%
\providecommand \urlprefix  [0]{URL }%
\providecommand \Eprint [0]{\href }%
\providecommand \doibase [0]{https://doi.org/}%
\providecommand \selectlanguage [0]{\@gobble}%
\providecommand \bibinfo  [0]{\@secondoftwo}%
\providecommand \bibfield  [0]{\@secondoftwo}%
\providecommand \translation [1]{[#1]}%
\providecommand \BibitemOpen [0]{}%
\providecommand \bibitemStop [0]{}%
\providecommand \bibitemNoStop [0]{.\EOS\space}%
\providecommand \EOS [0]{\spacefactor3000\relax}%
\providecommand \BibitemShut  [1]{\csname bibitem#1\endcsname}%
\let\auto@bib@innerbib\@empty

\section*{REFERENCES}

\bibitem [{\citenamefont {Moths}\ and\ \citenamefont
  {Witten}(2013)}]{moths2013full}%
  \BibitemOpen
  \bibfield  {author} {\bibinfo {author} {\bibfnamefont {B.}~\bibnamefont
  {Moths}}\ and\ \bibinfo {author} {\bibfnamefont {T.}~\bibnamefont {Witten}},\
  }\bibfield  {title} {\bibinfo {title} {Full alignment of colloidal objects by
  programmed forcing},\ }\href@noop {} {\bibfield  {journal} {\bibinfo
  {journal} {Physical Review Letters}\ }\textbf {\bibinfo {volume} {110}},\
  \bibinfo {pages} {028301} (\bibinfo {year} {2013})}\BibitemShut {NoStop}%
\bibitem [{\citenamefont {Björnmalm}\ \emph {et~al.}(2016)\citenamefont
  {Björnmalm}, \citenamefont {Faria}, \citenamefont {Chen}, \citenamefont
  {Cui},\ and\ \citenamefont {Caruso}}]{bjoornmalm2016dynamic}%
  \BibitemOpen
  \bibfield  {author} {\bibinfo {author} {\bibfnamefont {M.}~\bibnamefont
  {Björnmalm}}, \bibinfo {author} {\bibfnamefont {M.}~\bibnamefont {Faria}},
  \bibinfo {author} {\bibfnamefont {X.}~\bibnamefont {Chen}}, \bibinfo {author}
  {\bibfnamefont {J.}~\bibnamefont {Cui}},\ and\ \bibinfo {author}
  {\bibfnamefont {F.}~\bibnamefont {Caruso}},\ }\bibfield  {title} {\bibinfo
  {title} {Dynamic flow impacts cell--particle interactions: sedimentation and
  particle shape effects},\ }\href@noop {} {\bibfield  {journal} {\bibinfo
  {journal} {Langmuir}\ }\textbf {\bibinfo {volume} {32}},\ \bibinfo {pages}
  {10995} (\bibinfo {year} {2016})}\BibitemShut {NoStop}%
\bibitem [{\citenamefont {Saintillan}\ \emph {et~al.}(2005)\citenamefont
  {Saintillan}, \citenamefont {Darve},\ and\ \citenamefont
  {Shaqfeh}}]{saintillan2005smooth}%
  \BibitemOpen
  \bibfield  {author} {\bibinfo {author} {\bibfnamefont {D.}~\bibnamefont
  {Saintillan}}, \bibinfo {author} {\bibfnamefont {E.}~\bibnamefont {Darve}},\
  and\ \bibinfo {author} {\bibfnamefont {E.~S.}\ \bibnamefont {Shaqfeh}},\
  }\bibfield  {title} {\bibinfo {title} {A smooth particle-mesh {Ewald}
  algorithm for stokes suspension simulations: The sedimentation of fibers},\
  }\href@noop {} {\bibfield  {journal} {\bibinfo  {journal} {Physics of
  Fluids}\ }\textbf {\bibinfo {volume} {17}} (\bibinfo {year}
  {2005})}\BibitemShut {NoStop}%
\bibitem [{\citenamefont {Gruziel}\ \emph {et~al.}(2018)\citenamefont
  {Gruziel}, \citenamefont {Thyagarajan}, \citenamefont {Dietler},
  \citenamefont {Stasiak}, \citenamefont {Ekiel-Je{\.z}ewska},\ and\
  \citenamefont {Szymczak}}]{gruziel2018periodic}%
  \BibitemOpen
  \bibfield  {author} {\bibinfo {author} {\bibfnamefont {M.}~\bibnamefont
  {Gruziel}}, \bibinfo {author} {\bibfnamefont {K.}~\bibnamefont
  {Thyagarajan}}, \bibinfo {author} {\bibfnamefont {G.}~\bibnamefont
  {Dietler}}, \bibinfo {author} {\bibfnamefont {A.}~\bibnamefont {Stasiak}},
  \bibinfo {author} {\bibfnamefont {M.~L.}\ \bibnamefont
  {Ekiel-Je{\.z}ewska}},\ and\ \bibinfo {author} {\bibfnamefont
  {P.}~\bibnamefont {Szymczak}},\ }\bibfield  {title} {\bibinfo {title}
  {Periodic motion of sedimenting flexible knots},\ }\href@noop {} {\bibfield
  {journal} {\bibinfo  {journal} {Physical Review Letters}\ }\textbf {\bibinfo
  {volume} {121}},\ \bibinfo {pages} {127801} (\bibinfo {year}
  {2018})}\BibitemShut {NoStop}%
\bibitem [{\citenamefont {Vologodskii}\ \emph {et~al.}(1998)\citenamefont
  {Vologodskii}, \citenamefont {Crisona}, \citenamefont {Laurie}, \citenamefont
  {Pieranski}, \citenamefont {Katritch}, \citenamefont {Dubochet},\ and\
  \citenamefont {Stasiak}}]{vologodskii1998sedimentation}%
  \BibitemOpen
  \bibfield  {author} {\bibinfo {author} {\bibfnamefont {A.~V.}\ \bibnamefont
  {Vologodskii}}, \bibinfo {author} {\bibfnamefont {N.~J.}\ \bibnamefont
  {Crisona}}, \bibinfo {author} {\bibfnamefont {B.}~\bibnamefont {Laurie}},
  \bibinfo {author} {\bibfnamefont {P.}~\bibnamefont {Pieranski}}, \bibinfo
  {author} {\bibfnamefont {V.}~\bibnamefont {Katritch}}, \bibinfo {author}
  {\bibfnamefont {J.}~\bibnamefont {Dubochet}},\ and\ \bibinfo {author}
  {\bibfnamefont {A.}~\bibnamefont {Stasiak}},\ }\href@noop {} {\bibinfo
  {title} {Sedimentation and electrophoretic migration of {DNA} knots and
  catenanes}} (\bibinfo {year} {1998})\BibitemShut {NoStop}%
\bibitem [{\citenamefont {Weber}\ \emph {et~al.}(2013)\citenamefont {Weber},
  \citenamefont {Carlen}, \citenamefont {Dietler}, \citenamefont {Rawdon},\
  and\ \citenamefont {Stasiak}}]{weber2013sedimentation}%
  \BibitemOpen
  \bibfield  {author} {\bibinfo {author} {\bibfnamefont {C.}~\bibnamefont
  {Weber}}, \bibinfo {author} {\bibfnamefont {M.}~\bibnamefont {Carlen}},
  \bibinfo {author} {\bibfnamefont {G.}~\bibnamefont {Dietler}}, \bibinfo
  {author} {\bibfnamefont {E.~J.}\ \bibnamefont {Rawdon}},\ and\ \bibinfo
  {author} {\bibfnamefont {A.}~\bibnamefont {Stasiak}},\ }\bibfield  {title}
  {\bibinfo {title} {Sedimentation of macroscopic rigid knots and its relation
  to gel electrophoretic mobility of {DNA} knots},\ }\href@noop {} {\bibfield
  {journal} {\bibinfo  {journal} {Scientific Reports}\ }\textbf {\bibinfo
  {volume} {3}},\ \bibinfo {pages} {1091} (\bibinfo {year} {2013})}\BibitemShut
  {NoStop}%
\bibitem [{\citenamefont {Backes}\ \emph {et~al.}(2016)\citenamefont {Backes},
  \citenamefont {Hanlon}, \citenamefont {Szydlowska}, \citenamefont {Harvey},
  \citenamefont {Smith}, \citenamefont {Higgins},\ and\ \citenamefont
  {Coleman}}]{backes2016preparation}%
  \BibitemOpen
  \bibfield  {author} {\bibinfo {author} {\bibfnamefont {C.}~\bibnamefont
  {Backes}}, \bibinfo {author} {\bibfnamefont {D.}~\bibnamefont {Hanlon}},
  \bibinfo {author} {\bibfnamefont {B.~M.}\ \bibnamefont {Szydlowska}},
  \bibinfo {author} {\bibfnamefont {A.}~\bibnamefont {Harvey}}, \bibinfo
  {author} {\bibfnamefont {R.~J.}\ \bibnamefont {Smith}}, \bibinfo {author}
  {\bibfnamefont {T.~M.}\ \bibnamefont {Higgins}},\ and\ \bibinfo {author}
  {\bibfnamefont {J.~N.}\ \bibnamefont {Coleman}},\ }\bibfield  {title}
  {\bibinfo {title} {Preparation of liquid-exfoliated transition metal
  dichalcogenide nanosheets with controlled size and thickness: a state of the
  art protocol},\ }\href@noop {} {\bibfield  {journal} {\bibinfo  {journal}
  {JoVE (Journal of Visualized Experiments)}\ ,\ \bibinfo {pages} {e54806}}
  (\bibinfo {year} {2016})}\BibitemShut {NoStop}%
\bibitem [{\citenamefont {Xu}\ \emph {et~al.}(2018)\citenamefont {Xu},
  \citenamefont {Cao}, \citenamefont {Xue}, \citenamefont {Li},\ and\
  \citenamefont {Cai}}]{xu2018liquid}%
  \BibitemOpen
  \bibfield  {author} {\bibinfo {author} {\bibfnamefont {Y.}~\bibnamefont
  {Xu}}, \bibinfo {author} {\bibfnamefont {H.}~\bibnamefont {Cao}}, \bibinfo
  {author} {\bibfnamefont {Y.}~\bibnamefont {Xue}}, \bibinfo {author}
  {\bibfnamefont {B.}~\bibnamefont {Li}},\ and\ \bibinfo {author}
  {\bibfnamefont {W.}~\bibnamefont {Cai}},\ }\bibfield  {title} {\bibinfo
  {title} {Liquid-phase exfoliation of graphene: an overview on exfoliation
  media, techniques, and challenges},\ }\href@noop {} {\bibfield  {journal}
  {\bibinfo  {journal} {Nanomaterials}\ }\textbf {\bibinfo {volume} {8}},\
  \bibinfo {pages} {942} (\bibinfo {year} {2018})}\BibitemShut {NoStop}%
\bibitem [{\citenamefont {Zamani}\ \emph {et~al.}(2021)\citenamefont {Zamani},
  \citenamefont {Won}, \citenamefont {Salim}, \citenamefont {AlAmer},
  \citenamefont {Chang}, \citenamefont {Kumar}, \citenamefont {Amponsah},
  \citenamefont {Lim},\ and\ \citenamefont {Joo}}]{zamani2021ultralight}%
  \BibitemOpen
  \bibfield  {author} {\bibinfo {author} {\bibfnamefont {S.}~\bibnamefont
  {Zamani}}, \bibinfo {author} {\bibfnamefont {J.~S.}\ \bibnamefont {Won}},
  \bibinfo {author} {\bibfnamefont {M.}~\bibnamefont {Salim}}, \bibinfo
  {author} {\bibfnamefont {M.}~\bibnamefont {AlAmer}}, \bibinfo {author}
  {\bibfnamefont {C.-w.}\ \bibnamefont {Chang}}, \bibinfo {author}
  {\bibfnamefont {P.}~\bibnamefont {Kumar}}, \bibinfo {author} {\bibfnamefont
  {K.}~\bibnamefont {Amponsah}}, \bibinfo {author} {\bibfnamefont {A.~R.}\
  \bibnamefont {Lim}},\ and\ \bibinfo {author} {\bibfnamefont {Y.~L.}\
  \bibnamefont {Joo}},\ }\bibfield  {title} {\bibinfo {title} {Ultralight
  graphene/graphite hybrid fibers via entirely water-based processes and their
  application to density-controlled, high performance composites},\ }\href@noop
  {} {\bibfield  {journal} {\bibinfo  {journal} {Carbon}\ }\textbf {\bibinfo
  {volume} {173}},\ \bibinfo {pages} {880} (\bibinfo {year}
  {2021})}\BibitemShut {NoStop}%
\bibitem [{\citenamefont {Hernandez}\ \emph {et~al.}(2008)\citenamefont
  {Hernandez}, \citenamefont {Nicolosi}, \citenamefont {Lotya}, \citenamefont
  {Blighe}, \citenamefont {Sun}, \citenamefont {De}, \citenamefont {McGovern},
  \citenamefont {Holland}, \citenamefont {Byrne}, \citenamefont {Gun'Ko} \emph
  {et~al.}}]{hernandez2008high}%
  \BibitemOpen
  \bibfield  {author} {\bibinfo {author} {\bibfnamefont {Y.}~\bibnamefont
  {Hernandez}}, \bibinfo {author} {\bibfnamefont {V.}~\bibnamefont {Nicolosi}},
  \bibinfo {author} {\bibfnamefont {M.}~\bibnamefont {Lotya}}, \bibinfo
  {author} {\bibfnamefont {F.~M.}\ \bibnamefont {Blighe}}, \bibinfo {author}
  {\bibfnamefont {Z.}~\bibnamefont {Sun}}, \bibinfo {author} {\bibfnamefont
  {S.}~\bibnamefont {De}}, \bibinfo {author} {\bibfnamefont {I.~T.}\
  \bibnamefont {McGovern}}, \bibinfo {author} {\bibfnamefont {B.}~\bibnamefont
  {Holland}}, \bibinfo {author} {\bibfnamefont {M.}~\bibnamefont {Byrne}},
  \bibinfo {author} {\bibfnamefont {Y.~K.}\ \bibnamefont {Gun'Ko}}, \emph
  {et~al.},\ }\bibfield  {title} {\bibinfo {title} {High-yield production of
  graphene by liquid-phase exfoliation of graphite},\ }\href@noop {} {\bibfield
   {journal} {\bibinfo  {journal} {Nature Nanotechnology}\ }\textbf {\bibinfo
  {volume} {3}},\ \bibinfo {pages} {563} (\bibinfo {year} {2008})}\BibitemShut
  {NoStop}%
\bibitem [{\citenamefont {Wei}\ \emph {et~al.}(2014)\citenamefont {Wei},
  \citenamefont {Qiu}, \citenamefont {Ren}, \citenamefont {Zhang},
  \citenamefont {Wang},\ and\ \citenamefont {Weeks}}]{wei2014size}%
  \BibitemOpen
  \bibfield  {author} {\bibinfo {author} {\bibfnamefont {J.}~\bibnamefont
  {Wei}}, \bibinfo {author} {\bibfnamefont {J.}~\bibnamefont {Qiu}}, \bibinfo
  {author} {\bibfnamefont {L.}~\bibnamefont {Ren}}, \bibinfo {author}
  {\bibfnamefont {K.}~\bibnamefont {Zhang}}, \bibinfo {author} {\bibfnamefont
  {S.}~\bibnamefont {Wang}},\ and\ \bibinfo {author} {\bibfnamefont
  {B.}~\bibnamefont {Weeks}},\ }\bibfield  {title} {\bibinfo {title} {Size
  sorted multicolor fluorescence graphene oxide quantum dots obtained by
  differential velocity centrifugation},\ }\href@noop {} {\bibfield  {journal}
  {\bibinfo  {journal} {Science of Advanced Materials}\ }\textbf {\bibinfo
  {volume} {6}},\ \bibinfo {pages} {1052} (\bibinfo {year} {2014})}\BibitemShut
  {NoStop}%
\bibitem [{\citenamefont {Das}\ \emph {et~al.}(2018)\citenamefont {Das},
  \citenamefont {Bandyopadhyay},\ and\ \citenamefont
  {Pramanik}}]{das2018carbon}%
  \BibitemOpen
  \bibfield  {author} {\bibinfo {author} {\bibfnamefont {R.}~\bibnamefont
  {Das}}, \bibinfo {author} {\bibfnamefont {R.}~\bibnamefont {Bandyopadhyay}},\
  and\ \bibinfo {author} {\bibfnamefont {P.}~\bibnamefont {Pramanik}},\
  }\bibfield  {title} {\bibinfo {title} {Carbon quantum dots from natural
  resource: {A} review},\ }\href@noop {} {\bibfield  {journal} {\bibinfo
  {journal} {Materials Today Chemistry}\ }\textbf {\bibinfo {volume} {8}},\
  \bibinfo {pages} {96} (\bibinfo {year} {2018})}\BibitemShut {NoStop}%
\bibitem [{\citenamefont {Zhang}\ \emph {et~al.}(2018)\citenamefont {Zhang},
  \citenamefont {Xu}, \citenamefont {Chen}, \citenamefont {Gao},\ and\
  \citenamefont {Gao}}]{zhang2018large}%
  \BibitemOpen
  \bibfield  {author} {\bibinfo {author} {\bibfnamefont {P.}~\bibnamefont
  {Zhang}}, \bibinfo {author} {\bibfnamefont {B.}~\bibnamefont {Xu}}, \bibinfo
  {author} {\bibfnamefont {G.}~\bibnamefont {Chen}}, \bibinfo {author}
  {\bibfnamefont {C.}~\bibnamefont {Gao}},\ and\ \bibinfo {author}
  {\bibfnamefont {M.}~\bibnamefont {Gao}},\ }\bibfield  {title} {\bibinfo
  {title} {Large-scale synthesis of nitrogen doped mos2 quantum dots for
  efficient hydrogen evolution reaction},\ }\href@noop {} {\bibfield  {journal}
  {\bibinfo  {journal} {Electrochimica Acta}\ }\textbf {\bibinfo {volume}
  {270}},\ \bibinfo {pages} {256} (\bibinfo {year} {2018})}\BibitemShut
  {NoStop}%
\bibitem [{\citenamefont {Shashank}\ \emph {et~al.}(2023)\citenamefont
  {Shashank}, \citenamefont {Melikhov},\ and\ \citenamefont
  {Ekiel-Je{\.z}ewska}}]{shashank2023dynamics}%
  \BibitemOpen
  \bibfield  {author} {\bibinfo {author} {\bibfnamefont {H.}~\bibnamefont
  {Shashank}}, \bibinfo {author} {\bibfnamefont {Y.}~\bibnamefont {Melikhov}},\
  and\ \bibinfo {author} {\bibfnamefont {M.~L.}\ \bibnamefont
  {Ekiel-Je{\.z}ewska}},\ }\bibfield  {title} {\bibinfo {title} {Dynamics of
  ball chains and highly elastic fibres settling under gravity in a viscous
  fluid},\ }\href@noop {} {\bibfield  {journal} {\bibinfo  {journal} {Soft
  Matter}\ } (\bibinfo {year} {2023})}\BibitemShut {NoStop}%
\bibitem [{\citenamefont {Bukowicki}\ and\ \citenamefont
  {Ekiel-Je{\.z}ewska}(2019)}]{bukowicki2019sedimenting}%
  \BibitemOpen
  \bibfield  {author} {\bibinfo {author} {\bibfnamefont {M.}~\bibnamefont
  {Bukowicki}}\ and\ \bibinfo {author} {\bibfnamefont {M.~L.}\ \bibnamefont
  {Ekiel-Je{\.z}ewska}},\ }\bibfield  {title} {\bibinfo {title} {Sedimenting
  pairs of elastic microfilaments},\ }\href@noop {} {\bibfield  {journal}
  {\bibinfo  {journal} {Soft Matter}\ }\textbf {\bibinfo {volume} {15}},\
  \bibinfo {pages} {9405} (\bibinfo {year} {2019})}\BibitemShut {NoStop}%
\bibitem [{\citenamefont {Jianzhong}\ \emph {et~al.}(2003)\citenamefont
  {Jianzhong}, \citenamefont {Xing},\ and\ \citenamefont
  {Zhenjiang}}]{jianzhong2003effects}%
  \BibitemOpen
  \bibfield  {author} {\bibinfo {author} {\bibfnamefont {L.}~\bibnamefont
  {Jianzhong}}, \bibinfo {author} {\bibfnamefont {S.}~\bibnamefont {Xing}},\
  and\ \bibinfo {author} {\bibfnamefont {Y.}~\bibnamefont {Zhenjiang}},\
  }\bibfield  {title} {\bibinfo {title} {Effects of the aspect ratio on the
  sedimentation of a fiber in {Newtonian} fluids},\ }\href@noop {} {\bibfield
  {journal} {\bibinfo  {journal} {Journal of Aerosol Science}\ }\textbf
  {\bibinfo {volume} {34}},\ \bibinfo {pages} {909} (\bibinfo {year}
  {2003})}\BibitemShut {NoStop}%
\bibitem [{\citenamefont {Li}\ \emph {et~al.}(2013)\citenamefont {Li},
  \citenamefont {Manikantan}, \citenamefont {Saintillan},\ and\ \citenamefont
  {Spagnolie}}]{li2013sedimentation}%
  \BibitemOpen
  \bibfield  {author} {\bibinfo {author} {\bibfnamefont {L.}~\bibnamefont
  {Li}}, \bibinfo {author} {\bibfnamefont {H.}~\bibnamefont {Manikantan}},
  \bibinfo {author} {\bibfnamefont {D.}~\bibnamefont {Saintillan}},\ and\
  \bibinfo {author} {\bibfnamefont {S.~E.}\ \bibnamefont {Spagnolie}},\
  }\bibfield  {title} {\bibinfo {title} {The sedimentation of flexible
  filaments},\ }\href@noop {} {\bibfield  {journal} {\bibinfo  {journal}
  {Journal of Fluid Mechanics}\ }\textbf {\bibinfo {volume} {735}},\ \bibinfo
  {pages} {705} (\bibinfo {year} {2013})}\BibitemShut {NoStop}%
\bibitem [{\citenamefont {Lagomarsino}\ \emph {et~al.}(2005)\citenamefont
  {Lagomarsino}, \citenamefont {Pagonabarraga},\ and\ \citenamefont
  {Lowe}}]{lagomarsino2005hydrodynamic}%
  \BibitemOpen
  \bibfield  {author} {\bibinfo {author} {\bibfnamefont {M.~C.}\ \bibnamefont
  {Lagomarsino}}, \bibinfo {author} {\bibfnamefont {I.}~\bibnamefont
  {Pagonabarraga}},\ and\ \bibinfo {author} {\bibfnamefont {C.}~\bibnamefont
  {Lowe}},\ }\bibfield  {title} {\bibinfo {title} {Hydrodynamic induced
  deformation and orientation of a microscopic elastic filament},\ }\href@noop
  {} {\bibfield  {journal} {\bibinfo  {journal} {Physical Review Letters}\
  }\textbf {\bibinfo {volume} {94}},\ \bibinfo {pages} {148104} (\bibinfo
  {year} {2005})}\BibitemShut {NoStop}%
\bibitem [{\citenamefont {Saggiorato}\ \emph {et~al.}(2015)\citenamefont
  {Saggiorato}, \citenamefont {Elgeti}, \citenamefont {Winkler},\ and\
  \citenamefont {Gompper}}]{saggiorato2015conformations}%
  \BibitemOpen
  \bibfield  {author} {\bibinfo {author} {\bibfnamefont {G.}~\bibnamefont
  {Saggiorato}}, \bibinfo {author} {\bibfnamefont {J.}~\bibnamefont {Elgeti}},
  \bibinfo {author} {\bibfnamefont {R.~G.}\ \bibnamefont {Winkler}},\ and\
  \bibinfo {author} {\bibfnamefont {G.}~\bibnamefont {Gompper}},\ }\bibfield
  {title} {\bibinfo {title} {Conformations, hydrodynamic interactions, and
  instabilities of sedimenting semiflexible filaments},\ }\href@noop {}
  {\bibfield  {journal} {\bibinfo  {journal} {Soft Matter}\ }\textbf {\bibinfo
  {volume} {11}},\ \bibinfo {pages} {7337} (\bibinfo {year}
  {2015})}\BibitemShut {NoStop}%
\bibitem [{\citenamefont {Cunha}\ \emph {et~al.}(2022)\citenamefont {Cunha},
  \citenamefont {Zhao}, \citenamefont {MacKintosh},\ and\ \citenamefont
  {Biswal}}]{cunha2022settling}%
  \BibitemOpen
  \bibfield  {author} {\bibinfo {author} {\bibfnamefont {L.~H.}\ \bibnamefont
  {Cunha}}, \bibinfo {author} {\bibfnamefont {J.}~\bibnamefont {Zhao}},
  \bibinfo {author} {\bibfnamefont {F.~C.}\ \bibnamefont {MacKintosh}},\ and\
  \bibinfo {author} {\bibfnamefont {S.~L.}\ \bibnamefont {Biswal}},\ }\bibfield
   {title} {\bibinfo {title} {Settling dynamics of {Brownian} chains in viscous
  fluids},\ }\href@noop {} {\bibfield  {journal} {\bibinfo  {journal} {Physical
  Review Fluids}\ }\textbf {\bibinfo {volume} {7}},\ \bibinfo {pages} {034303}
  (\bibinfo {year} {2022})}\BibitemShut {NoStop}%
\bibitem [{\citenamefont {Gruziel-S{\l}omka}\ \emph {et~al.}(2019)\citenamefont
  {Gruziel-S{\l}omka}, \citenamefont {Kondratiuk}, \citenamefont {Szymczak},\
  and\ \citenamefont {Ekiel-Je{\.z}ewska}}]{gruziel2019stokesian}%
  \BibitemOpen
  \bibfield  {author} {\bibinfo {author} {\bibfnamefont {M.}~\bibnamefont
  {Gruziel-S{\l}omka}}, \bibinfo {author} {\bibfnamefont {P.}~\bibnamefont
  {Kondratiuk}}, \bibinfo {author} {\bibfnamefont {P.}~\bibnamefont
  {Szymczak}},\ and\ \bibinfo {author} {\bibfnamefont {M.~L.}\ \bibnamefont
  {Ekiel-Je{\.z}ewska}},\ }\bibfield  {title} {\bibinfo {title} {Stokesian
  dynamics of sedimenting elastic rings},\ }\href@noop {} {\bibfield  {journal}
  {\bibinfo  {journal} {Soft Matter}\ }\textbf {\bibinfo {volume} {15}},\
  \bibinfo {pages} {7262} (\bibinfo {year} {2019})}\BibitemShut {NoStop}%
\bibitem [{\citenamefont {Makino}\ and\ \citenamefont
  {Doi}(2003)}]{makino2003sedimentation}%
  \BibitemOpen
  \bibfield  {author} {\bibinfo {author} {\bibfnamefont {M.}~\bibnamefont
  {Makino}}\ and\ \bibinfo {author} {\bibfnamefont {M.}~\bibnamefont {Doi}},\
  }\bibfield  {title} {\bibinfo {title} {Sedimentation of a particle with
  translation--rotation coupling},\ }\href@noop {} {\bibfield  {journal}
  {\bibinfo  {journal} {Journal of the Physical Society of Japan}\ }\textbf
  {\bibinfo {volume} {72}},\ \bibinfo {pages} {2699} (\bibinfo {year}
  {2003})}\BibitemShut {NoStop}%
\bibitem [{\citenamefont {Chajwa}\ \emph {et~al.}(2019)\citenamefont {Chajwa},
  \citenamefont {Menon},\ and\ \citenamefont {Ramaswamy}}]{chajwa2019kepler}%
  \BibitemOpen
  \bibfield  {author} {\bibinfo {author} {\bibfnamefont {R.}~\bibnamefont
  {Chajwa}}, \bibinfo {author} {\bibfnamefont {N.}~\bibnamefont {Menon}},\ and\
  \bibinfo {author} {\bibfnamefont {S.}~\bibnamefont {Ramaswamy}},\ }\bibfield
  {title} {\bibinfo {title} {Kepler orbits in pairs of disks settling in a
  viscous fluid},\ }\href@noop {} {\bibfield  {journal} {\bibinfo  {journal}
  {Physical Review Letters}\ }\textbf {\bibinfo {volume} {122}},\ \bibinfo
  {pages} {224501} (\bibinfo {year} {2019})}\BibitemShut {NoStop}%
\bibitem [{\citenamefont {Salez}\ and\ \citenamefont
  {Mahadevan}(2015)}]{salez2015elastohydrodynamics}%
  \BibitemOpen
  \bibfield  {author} {\bibinfo {author} {\bibfnamefont {T.}~\bibnamefont
  {Salez}}\ and\ \bibinfo {author} {\bibfnamefont {L.}~\bibnamefont
  {Mahadevan}},\ }\bibfield  {title} {\bibinfo {title} {Elastohydrodynamics of
  a sliding, spinning and sedimenting cylinder near a soft wall},\ }\href@noop
  {} {\bibfield  {journal} {\bibinfo  {journal} {Journal of Fluid Mechanics}\
  }\textbf {\bibinfo {volume} {779}},\ \bibinfo {pages} {181} (\bibinfo {year}
  {2015})}\BibitemShut {NoStop}%
\bibitem [{\citenamefont {Saintyves}\ \emph {et~al.}(2016)\citenamefont
  {Saintyves}, \citenamefont {Jules}, \citenamefont {Salez},\ and\
  \citenamefont {Mahadevan}}]{saintyves2016self}%
  \BibitemOpen
  \bibfield  {author} {\bibinfo {author} {\bibfnamefont {B.}~\bibnamefont
  {Saintyves}}, \bibinfo {author} {\bibfnamefont {T.}~\bibnamefont {Jules}},
  \bibinfo {author} {\bibfnamefont {T.}~\bibnamefont {Salez}},\ and\ \bibinfo
  {author} {\bibfnamefont {L.}~\bibnamefont {Mahadevan}},\ }\bibfield  {title}
  {\bibinfo {title} {Self-sustained lift and low friction via soft
  lubrication},\ }\href@noop {} {\bibfield  {journal} {\bibinfo  {journal}
  {Proceedings of the National Academy of Sciences}\ }\textbf {\bibinfo
  {volume} {113}},\ \bibinfo {pages} {5847} (\bibinfo {year}
  {2016})}\BibitemShut {NoStop}%
\bibitem [{\citenamefont {Rallabandi}\ \emph {et~al.}(2017)\citenamefont
  {Rallabandi}, \citenamefont {Saintyves}, \citenamefont {Jules}, \citenamefont
  {Salez}, \citenamefont {Sch{\"o}necker}, \citenamefont {Mahadevan},\ and\
  \citenamefont {Stone}}]{rallabandi2017rotation}%
  \BibitemOpen
  \bibfield  {author} {\bibinfo {author} {\bibfnamefont {B.}~\bibnamefont
  {Rallabandi}}, \bibinfo {author} {\bibfnamefont {B.}~\bibnamefont
  {Saintyves}}, \bibinfo {author} {\bibfnamefont {T.}~\bibnamefont {Jules}},
  \bibinfo {author} {\bibfnamefont {T.}~\bibnamefont {Salez}}, \bibinfo
  {author} {\bibfnamefont {C.}~\bibnamefont {Sch{\"o}necker}}, \bibinfo
  {author} {\bibfnamefont {L.}~\bibnamefont {Mahadevan}},\ and\ \bibinfo
  {author} {\bibfnamefont {H.~A.}\ \bibnamefont {Stone}},\ }\bibfield  {title}
  {\bibinfo {title} {Rotation of an immersed cylinder sliding near a thin
  elastic coating},\ }\href@noop {} {\bibfield  {journal} {\bibinfo  {journal}
  {Physical Review Fluids}\ }\textbf {\bibinfo {volume} {2}},\ \bibinfo {pages}
  {074102} (\bibinfo {year} {2017})}\BibitemShut {NoStop}%
\bibitem [{\citenamefont {Mitchell}\ and\ \citenamefont
  {Spagnolie}(2015)}]{mitchell2015sedimentation}%
  \BibitemOpen
  \bibfield  {author} {\bibinfo {author} {\bibfnamefont {W.~H.}\ \bibnamefont
  {Mitchell}}\ and\ \bibinfo {author} {\bibfnamefont {S.~E.}\ \bibnamefont
  {Spagnolie}},\ }\bibfield  {title} {\bibinfo {title} {Sedimentation of
  spheroidal bodies near walls in viscous fluids: glancing, reversing, tumbling
  and sliding},\ }\href@noop {} {\bibfield  {journal} {\bibinfo  {journal}
  {Journal of Fluid Mechanics}\ }\textbf {\bibinfo {volume} {772}},\ \bibinfo
  {pages} {600} (\bibinfo {year} {2015})}\BibitemShut {NoStop}%
\bibitem [{\citenamefont {Peltom{\"a}ki}\ and\ \citenamefont
  {Gompper}(2013)}]{peltomaki2013sedimentation}%
  \BibitemOpen
  \bibfield  {author} {\bibinfo {author} {\bibfnamefont {M.}~\bibnamefont
  {Peltom{\"a}ki}}\ and\ \bibinfo {author} {\bibfnamefont {G.}~\bibnamefont
  {Gompper}},\ }\bibfield  {title} {\bibinfo {title} {Sedimentation of single
  red blood cells},\ }\href@noop {} {\bibfield  {journal} {\bibinfo  {journal}
  {Soft Matter}\ }\textbf {\bibinfo {volume} {9}},\ \bibinfo {pages} {8346}
  (\bibinfo {year} {2013})}\BibitemShut {NoStop}%
\bibitem [{\citenamefont {Matsunaga}\ \emph {et~al.}(2016)\citenamefont
  {Matsunaga}, \citenamefont {Imai}, \citenamefont {Wagner},\ and\
  \citenamefont {Ishikawa}}]{matsunaga2016reorientation}%
  \BibitemOpen
  \bibfield  {author} {\bibinfo {author} {\bibfnamefont {D.}~\bibnamefont
  {Matsunaga}}, \bibinfo {author} {\bibfnamefont {Y.}~\bibnamefont {Imai}},
  \bibinfo {author} {\bibfnamefont {C.}~\bibnamefont {Wagner}},\ and\ \bibinfo
  {author} {\bibfnamefont {T.}~\bibnamefont {Ishikawa}},\ }\bibfield  {title}
  {\bibinfo {title} {Reorientation of a single red blood cell during
  sedimentation},\ }\href@noop {} {\bibfield  {journal} {\bibinfo  {journal}
  {Journal of Fluid Mechanics}\ }\textbf {\bibinfo {volume} {806}},\ \bibinfo
  {pages} {102} (\bibinfo {year} {2016})}\BibitemShut {NoStop}%
\bibitem [{\citenamefont {Becker}\ \emph {et~al.}(1996)\citenamefont {Becker},
  \citenamefont {McKinley},\ and\ \citenamefont
  {Stone}}]{becker1996sedimentation}%
  \BibitemOpen
  \bibfield  {author} {\bibinfo {author} {\bibfnamefont {L.}~\bibnamefont
  {Becker}}, \bibinfo {author} {\bibfnamefont {G.}~\bibnamefont {McKinley}},\
  and\ \bibinfo {author} {\bibfnamefont {H.}~\bibnamefont {Stone}},\ }\bibfield
   {title} {\bibinfo {title} {Sedimentation of a sphere near a plane wall: weak
  non-{Newtonian} and inertial effects},\ }\href@noop {} {\bibfield  {journal}
  {\bibinfo  {journal} {Journal of Non-Newtonian Fluid Mechanics}\ }\textbf
  {\bibinfo {volume} {63}},\ \bibinfo {pages} {201} (\bibinfo {year}
  {1996})}\BibitemShut {NoStop}%
\bibitem [{\citenamefont {Singh}\ and\ \citenamefont
  {Joseph}(2000)}]{singh2000sedimentation}%
  \BibitemOpen
  \bibfield  {author} {\bibinfo {author} {\bibfnamefont {P.}~\bibnamefont
  {Singh}}\ and\ \bibinfo {author} {\bibfnamefont {D.}~\bibnamefont {Joseph}},\
  }\bibfield  {title} {\bibinfo {title} {Sedimentation of a sphere near a
  vertical wall in an {Oldroyd-B} fluid},\ }\href@noop {} {\bibfield  {journal}
  {\bibinfo  {journal} {Journal of Non-Newtonian Fluid Mechanics}\ }\textbf
  {\bibinfo {volume} {94}},\ \bibinfo {pages} {179} (\bibinfo {year}
  {2000})}\BibitemShut {NoStop}%
\bibitem [{\citenamefont {Graham}(2018)}]{graham2018microhydrodynamics}%
  \BibitemOpen
  \bibfield  {author} {\bibinfo {author} {\bibfnamefont {M.~D.}\ \bibnamefont
  {Graham}},\ }\href@noop {} {\emph {\bibinfo {title} {Microhydrodynamics,
  {Brownian} Motion, and {Complex Fluids}}}},\ Vol.~\bibinfo {volume} {58}\
  (\bibinfo  {publisher} {Cambridge University Press},\ \bibinfo {year}
  {2018})\BibitemShut {NoStop}%
\bibitem [{\citenamefont {Russel}\ \emph {et~al.}(1977)\citenamefont {Russel},
  \citenamefont {Hinch}, \citenamefont {Leal},\ and\ \citenamefont
  {Tieffenbruck}}]{russel1977rods}%
  \BibitemOpen
  \bibfield  {author} {\bibinfo {author} {\bibfnamefont {W.}~\bibnamefont
  {Russel}}, \bibinfo {author} {\bibfnamefont {E.}~\bibnamefont {Hinch}},
  \bibinfo {author} {\bibfnamefont {L.~G.}\ \bibnamefont {Leal}},\ and\
  \bibinfo {author} {\bibfnamefont {G.}~\bibnamefont {Tieffenbruck}},\
  }\bibfield  {title} {\bibinfo {title} {Rods falling near a vertical wall},\
  }\href@noop {} {\bibfield  {journal} {\bibinfo  {journal} {Journal of Fluid
  Mechanics}\ }\textbf {\bibinfo {volume} {83}},\ \bibinfo {pages} {273}
  (\bibinfo {year} {1977})}\BibitemShut {NoStop}%
\bibitem [{\citenamefont {Xu}\ and\ \citenamefont
  {Green}(2014)}]{xu2014brownian}%
  \BibitemOpen
  \bibfield  {author} {\bibinfo {author} {\bibfnamefont {Y.}~\bibnamefont
  {Xu}}\ and\ \bibinfo {author} {\bibfnamefont {M.~J.}\ \bibnamefont {Green}},\
  }\bibfield  {title} {\bibinfo {title} {Brownian dynamics simulations of
  nanosheet solutions under shear},\ }\href@noop {} {\bibfield  {journal}
  {\bibinfo  {journal} {The Journal of Chemical Physics}\ }\textbf {\bibinfo
  {volume} {141}} (\bibinfo {year} {2014})}\BibitemShut {NoStop}%
\bibitem [{\citenamefont {Silmore}\ \emph {et~al.}(2021)\citenamefont
  {Silmore}, \citenamefont {Strano},\ and\ \citenamefont
  {Swan}}]{silmore2021buckling}%
  \BibitemOpen
  \bibfield  {author} {\bibinfo {author} {\bibfnamefont {K.~S.}\ \bibnamefont
  {Silmore}}, \bibinfo {author} {\bibfnamefont {M.~S.}\ \bibnamefont
  {Strano}},\ and\ \bibinfo {author} {\bibfnamefont {J.~W.}\ \bibnamefont
  {Swan}},\ }\bibfield  {title} {\bibinfo {title} {Buckling, crumpling, and
  tumbling of semiflexible sheets in simple shear flow},\ }\href@noop {}
  {\bibfield  {journal} {\bibinfo  {journal} {Soft Matter}\ }\textbf {\bibinfo
  {volume} {17}},\ \bibinfo {pages} {4707} (\bibinfo {year}
  {2021})}\BibitemShut {NoStop}%
\bibitem [{\citenamefont {Perrin}\ \emph {et~al.}(2023)\citenamefont {Perrin},
  \citenamefont {Li},\ and\ \citenamefont {Botto}}]{perrin2023hydrodynamic}%
  \BibitemOpen
  \bibfield  {author} {\bibinfo {author} {\bibfnamefont {H.}~\bibnamefont
  {Perrin}}, \bibinfo {author} {\bibfnamefont {H.}~\bibnamefont {Li}},\ and\
  \bibinfo {author} {\bibfnamefont {L.}~\bibnamefont {Botto}},\ }\bibfield
  {title} {\bibinfo {title} {Hydrodynamic interactions change the buckling
  threshold of parallel flexible sheets in shear flow},\ }\href@noop {}
  {\bibfield  {journal} {\bibinfo  {journal} {arXiv preprint arXiv:2303.05282}\
  } (\bibinfo {year} {2023})}\BibitemShut {NoStop}%
\bibitem [{\citenamefont {Yu}\ and\ \citenamefont {Graham}(2021)}]{yu2021coil}%
  \BibitemOpen
  \bibfield  {author} {\bibinfo {author} {\bibfnamefont {Y.}~\bibnamefont
  {Yu}}\ and\ \bibinfo {author} {\bibfnamefont {M.~D.}\ \bibnamefont
  {Graham}},\ }\bibfield  {title} {\bibinfo {title} {Coil--stretch-like
  transition of elastic sheets in extensional flows},\ }\href@noop {}
  {\bibfield  {journal} {\bibinfo  {journal} {Soft Matter}\ }\textbf {\bibinfo
  {volume} {17}},\ \bibinfo {pages} {543} (\bibinfo {year} {2021})}\BibitemShut
  {NoStop}%
\bibitem [{\citenamefont {Yu}\ and\ \citenamefont
  {Graham}(2022)}]{yu2022wrinkling}%
  \BibitemOpen
  \bibfield  {author} {\bibinfo {author} {\bibfnamefont {Y.}~\bibnamefont
  {Yu}}\ and\ \bibinfo {author} {\bibfnamefont {M.~D.}\ \bibnamefont
  {Graham}},\ }\bibfield  {title} {\bibinfo {title} {Wrinkling and multiplicity
  in the dynamics of deformable sheets in uniaxial extensional flow},\
  }\href@noop {} {\bibfield  {journal} {\bibinfo  {journal} {Physical Review
  Fluids}\ }\textbf {\bibinfo {volume} {7}},\ \bibinfo {pages} {023601}
  (\bibinfo {year} {2022})}\BibitemShut {NoStop}%
\bibitem [{\citenamefont {Fedosov}\ \emph {et~al.}(2010)\citenamefont
  {Fedosov}, \citenamefont {Caswell},\ and\ \citenamefont
  {Karniadakis}}]{Fedosov2010}%
  \BibitemOpen
  \bibfield  {author} {\bibinfo {author} {\bibfnamefont {D.~A.}\ \bibnamefont
  {Fedosov}}, \bibinfo {author} {\bibfnamefont {B.}~\bibnamefont {Caswell}},\
  and\ \bibinfo {author} {\bibfnamefont {G.~E.}\ \bibnamefont {Karniadakis}},\
  }\bibfield  {title} {\bibinfo {title} {Systematic coarse-graining of
  spectrin-level red blood cell models},\ }\href@noop {} {\bibfield  {journal}
  {\bibinfo  {journal} {Computer Methods in Applied Mechanics and Engineering}\
  }\textbf {\bibinfo {volume} {199}},\ \bibinfo {pages} {1937} (\bibinfo {year}
  {2010})}\BibitemShut {NoStop}%
\bibitem [{\citenamefont {Fedosov}(2010)}]{fedosov2010multiscale}%
  \BibitemOpen
  \bibfield  {author} {\bibinfo {author} {\bibfnamefont {D.~A.}\ \bibnamefont
  {Fedosov}},\ }\href@noop {} {\emph {\bibinfo {title} {Multiscale modeling of
  blood flow and soft matter}}}\ (\bibinfo  {publisher} {Citeseer},\ \bibinfo
  {year} {2010})\BibitemShut {NoStop}%
\bibitem [{\citenamefont {Charrier}\ \emph {et~al.}(1989)\citenamefont
  {Charrier}, \citenamefont {Shrivastava},\ and\ \citenamefont
  {Wu}}]{charrier1989free}%
  \BibitemOpen
  \bibfield  {author} {\bibinfo {author} {\bibfnamefont {J.}~\bibnamefont
  {Charrier}}, \bibinfo {author} {\bibfnamefont {S.~C.}\ \bibnamefont
  {Shrivastava}},\ and\ \bibinfo {author} {\bibfnamefont {R.}~\bibnamefont
  {Wu}},\ }\bibfield  {title} {\bibinfo {title} {Free and constrained inflation
  of elastic membranes in relation to thermoforming---non-axisymmetric
  problems},\ }\href@noop {} {\bibfield  {journal} {\bibinfo  {journal} {The
  Journal of Strain Analysis for Engineering Design}\ }\textbf {\bibinfo
  {volume} {24}},\ \bibinfo {pages} {55} (\bibinfo {year} {1989})}\BibitemShut
  {NoStop}%
\bibitem [{\citenamefont {Cortez}\ \emph {et~al.}(2005)\citenamefont {Cortez},
  \citenamefont {Fauci},\ and\ \citenamefont {Medovikov}}]{cortez2005method}%
  \BibitemOpen
  \bibfield  {author} {\bibinfo {author} {\bibfnamefont {R.}~\bibnamefont
  {Cortez}}, \bibinfo {author} {\bibfnamefont {L.}~\bibnamefont {Fauci}},\ and\
  \bibinfo {author} {\bibfnamefont {A.}~\bibnamefont {Medovikov}},\ }\bibfield
  {title} {\bibinfo {title} {The method of regularized {Stokeslets} in three
  dimensions: analysis, validation, and application to helical swimming},\
  }\href@noop {} {\bibfield  {journal} {\bibinfo  {journal} {Physics of
  Fluids}\ }\textbf {\bibinfo {volume} {17}},\ \bibinfo {pages} {031504}
  (\bibinfo {year} {2005})}\BibitemShut {NoStop}%
\bibitem [{\citenamefont {Blake}(1971)}]{blake1971note}%
  \BibitemOpen
  \bibfield  {author} {\bibinfo {author} {\bibfnamefont {J.~R.}\ \bibnamefont
  {Blake}},\ }\bibfield  {title} {\bibinfo {title} {A note on the image system
  for a {Stokeslet} in a no-slip boundary},\ }in\ \href@noop {} {\emph
  {\bibinfo {booktitle} {Mathematical Proceedings of the Cambridge
  Philosophical Society}}},\ Vol.~\bibinfo {volume} {70}\ (\bibinfo
  {organization} {Cambridge University Press},\ \bibinfo {year} {1971})\ pp.\
  \bibinfo {pages} {303--310}\BibitemShut {NoStop}%
\bibitem [{\citenamefont {Ainley}\ \emph {et~al.}(2008)\citenamefont {Ainley},
  \citenamefont {Durkin}, \citenamefont {Embid}, \citenamefont {Boindala},\
  and\ \citenamefont {Cortez}}]{ainley2008method}%
  \BibitemOpen
  \bibfield  {author} {\bibinfo {author} {\bibfnamefont {J.}~\bibnamefont
  {Ainley}}, \bibinfo {author} {\bibfnamefont {S.}~\bibnamefont {Durkin}},
  \bibinfo {author} {\bibfnamefont {R.}~\bibnamefont {Embid}}, \bibinfo
  {author} {\bibfnamefont {P.}~\bibnamefont {Boindala}},\ and\ \bibinfo
  {author} {\bibfnamefont {R.}~\bibnamefont {Cortez}},\ }\bibfield  {title}
  {\bibinfo {title} {The method of images for regularized {Stokeslets}},\
  }\href@noop {} {\bibfield  {journal} {\bibinfo  {journal} {Journal of
  Computational Physics}\ }\textbf {\bibinfo {volume} {227}},\ \bibinfo {pages}
  {4600} (\bibinfo {year} {2008})}\BibitemShut {NoStop}%
\bibitem [{\citenamefont {Cortez}\ and\ \citenamefont
  {Varela}(2015)}]{cortez2015general}%
  \BibitemOpen
  \bibfield  {author} {\bibinfo {author} {\bibfnamefont {R.}~\bibnamefont
  {Cortez}}\ and\ \bibinfo {author} {\bibfnamefont {D.}~\bibnamefont
  {Varela}},\ }\bibfield  {title} {\bibinfo {title} {A general system of images
  for regularized {Stokeslets} and other elements near a plane wall},\
  }\href@noop {} {\bibfield  {journal} {\bibinfo  {journal} {Journal of
  Computational Physics}\ }\textbf {\bibinfo {volume} {285}},\ \bibinfo {pages}
  {41} (\bibinfo {year} {2015})}\BibitemShut {NoStop}%
\end{thebibliography}

%

\end{document}